\documentclass[journal=jacsat ,manuscript=letter]{achemso}
\setkeys{acs}{maxauthors = 10}

\usepackage{hyperref}
\hypersetup{backref,colorlinks=true,}
\usepackage{xr}
\usepackage{chemformula} 
\usepackage[T1]{fontenc} 

\usepackage{graphicx}
\graphicspath{{./figures/}}
\usepackage{amsmath,amssymb}
\usepackage{float}
\usepackage{multirow}
\usepackage{tabularx}
\usepackage{pifont}
\usepackage{wrapfig}
\usepackage[caption=false]{subfig}
\usepackage{miller}
\usepackage{orcidlink}
\usepackage{todonotes}
\usepackage{makecell}

\mciteErrorOnUnknownfalse

\usepackage{tikz}
\usetikzlibrary{calc,arrows,shapes,positioning}
\usetikzlibrary{patterns,snakes}
\usetikzlibrary{positioning,decorations.pathreplacing}

\definecolor{sangria}{rgb}{0.57, 0.0, 0.04}
\definecolor{arsenic}{rgb}{0.23, 0.27, 0.29}
\definecolor{prussianblue}{rgb}{0.0, 0.19, 0.33}
\definecolor{phthalogreen}{rgb}{0.07, 0.21, 0.14}
\definecolor{dgreen}{rgb}{0.0, 0.4, 0.2}

\def\fig#1{Fig.~\ref{fig:#1}}
\def\eq#1{Eq.~\eqref{eq:#1}}
\def\tab#1{Table~\ref{tab:#1}}

\title{Dendrite Suppression in Zn Batteries Through Hetero-Epitaxial Residual Stresses Shield}
\author{Musanna Galib}
\affiliation{Department of Mechanical Engineering, The University of British Columbia, 2054 - 6250 Applied Science Lane, Vancouver, BC, V6T 1Z4, Canada}
\author{Amardeep Amardeep}
\affiliation{School of Engineering, Faculty of Applied Science, The University of British Columbia, Okanagan Campus, Kelowna, BC, V1V 1V7, Canada}
\author{Jian Liu}
\affiliation{School of Engineering, Faculty of Applied Science, The University of British Columbia, Okanagan Campus, Kelowna, BC, V1V 1V7, Canada}
\email{jian.liu@ubc.ca}
\author{Mauricio Ponga}
\affiliation{Department of Mechanical Engineering, The University of British Columbia, 2054 - 6250 Applied Science Lane, Vancouver, BC, V6T 1Z4, Canada}
\email{mponga@mech.ubc.ca}

\begin{document}

\newpage

\begin{abstract}
Dendrite formation is a long-standing problem for the commercial application of aqueous zinc (Zn)-ion batteries (AZIB). 
Here, we investigate the effect of hetero-epitaxial residual stresses due to layered coatings on dendrite suppression. We found that atomic and molecular layered coatings can substantially reduce dendritic growth in AZIB by providing shielding due to residual stresses, even at single and a few layers of coatings. Through a combined experimental and numerical approach, we demonstrate that the residual stresses developed due to the coating of the Zn anodes significantly reduced the chemical potential polarization around dendrite embryos, forcing the deposition of zinc in the regions adjacent to the protuberances. This, in turn, results in a slower rate of dendritic growth, and eventually, dendrite suppression. The fundamental understanding of the effect of residual stresses due to coatings demonstrated herein can be extended to various metal anode batteries such as Li or Na.

\end{abstract}
%

\begin{center}
\textbf{TOC Graphic}
\end{center}

\begin{figure}[H]  
    \centering
    \fbox{\includegraphics[width=3in, height=2in]{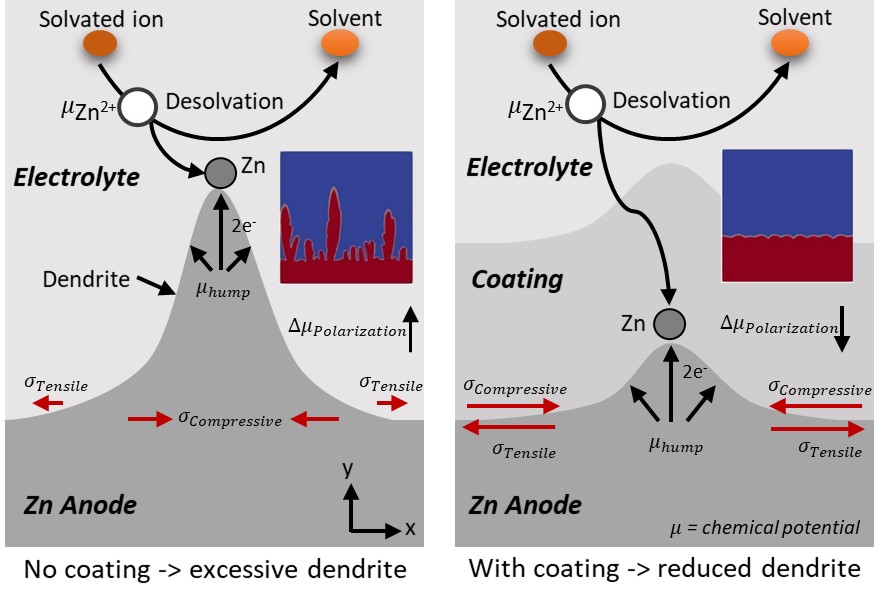}} 
\end{figure}

\maketitle

\newpage

\section{\label{sec:Introduction}Introduction}
\par Rechargeable metal anode batteries with, such as Zn~\cite{10.1016/j.joule.2018.11.007}, Na~\cite{10.1002/aenm.201702869},  Al~\cite{10.1038/nature14340}, Ca~\cite{10.1038/ncomms10999}, stand out as favorable contenders for the next-generation energy storage devices. Their attractiveness stems from their excellent gravimetric energy density, low redox potential, cost-effectiveness, and enhanced safety characteristics~\cite{10.1002/adfm.202004187}. However, a common obstacle hindering the widespread adoption of these metal anode-based batteries is the challenge posed by suboptimal cycling performance and safety concerns, specifically linked to dendrite growth~\cite{10.1021/acsenergylett.6b00650}. 
\par Zn metal offers notable advantages, including substantial capacity (820 mAh$\cdot$g\textsuperscript{-1} and 5848 Ah$\cdot$L\textsuperscript{-1} of Zn compared to 3861 mAh$\cdot$g\textsuperscript{-1} and 2061 Ah$\cdot$L\textsuperscript{-1} of Li)~\cite{10.1021/acs.chemrev.8b00422}, greater concentration in the earth's crust (70 ppm of Zn compared to 20 ppm of Li)~\cite{10.1134/S001670290601006X}, and cost-effectiveness (USD $1.8-4$ per kg of Zn compared to USD $5.8-80$ per kg of Li)~\cite{10.1038/s41467-024-48368-0}. These factors have spurred the current revival of rechargeable aqueous Zn ion batteries (AZIBs), known for their safety and affordability ($<$ USD 10 per kWh of AZIBs~\cite{10.1002/anie.201904174} compared to USD $135$ per kWh of LIBs)~\cite{10.1038/nenergy.2017.110}. 
However, akin to Li metal, anodic insufficiency remains a persistent challenge in AZIBs due to dendritic Zn growth, Zn passivation and corrosion, and H\textsubscript{2} evolution~\cite{10.1002/smll.202206634}. Zn dendrites, usually deemed as the primary reason for internal short-circuits in AZIBs, proliferate next to the separator, eventually breaching the separator. Identical to Li, the distal ends of Zn dendrites undergo passivation, resulting in ``dead'' Zn. This accumulation leads to electrolyte insufficiency, anodic capacity loss, increased cell resistance and polarization~\cite{10.1039/C7TA00371D}. Impurities in Zn give rise to by-products, such as $\mathrm{Zn(OH)_2}$ or $\mathrm{Zn_{4}SO_{4}(OH)_{6}.nH_{2}O}$, in electrolytes alongside gas evolution. These problems are rooted in the same source: an unfavorable interaction at the anode and electrolyte contact~\cite{10.1002/smll.202206634}.
\par The electroplating of Zn involves a sequence of events encompassing $\mathrm{Zn^{2+}}$ ion diffusion and migration, then reduction and nucleation, and finally, crystal growth. As a diffusion-controlled process, Zn nucleation is influenced by the applied electric field and, hence, ion distribution~\cite{10.1021/acsenergylett.0c02028}. The evolution of Zn dendrites is perpetuated by inhomogeneous nucleation due to the localized high electric field and preferential accumulation at the tips with large curvature radii~\cite{10.1002/adma.202001854}. However, the pivotal roles are played by electrochemical operational requirements such as current density and plating/stripping capacity in shaping the dendritic structure of Zn~\cite{10.1002/adma.201903778}. 
Recent studies have utilized various techniques to investigate crystal structures, for instance, examining nanoscale or microscale forms of dendrites through ex-situ electron microscopy and observing lateral morphological growth of the Zn anode via in-situ optical microscopy (OM)~\cite{10.1021/acsami.2c19895}. Furthermore, surface coatings could be an effective strategy to suppress dendrite growth in metallic anodes by providing nucleation uniformity~\cite{10.1016/j.nanoen.2023.108306,10.1002/advs.202100309,NL:AlCoating} .

\par Here, we demonstrate how residual stresses can suppress dendrite nucleation and growth by altering the chemical potential near the neighborhood of dendrite embryos. First, we performed an operando investigation of dendritic growth in $\mathrm{Zn||Zn}$ symmetric cells through in-situ OM to correlate the microstructure evolution and electrochemical conditions, specifically in thin films. 
 Further, to unveil the dendrite formation mechanics based on these experiments, a mesoscale simulation technique using phase-field modeling (PFM)~\cite{10.1021/acsenergylett.8b01009} was used to capture the dendrite evolution under various levels of residual stresses. 

\begin{figure}[h]
\centering
\label{}
\includegraphics[width=0.99\linewidth]{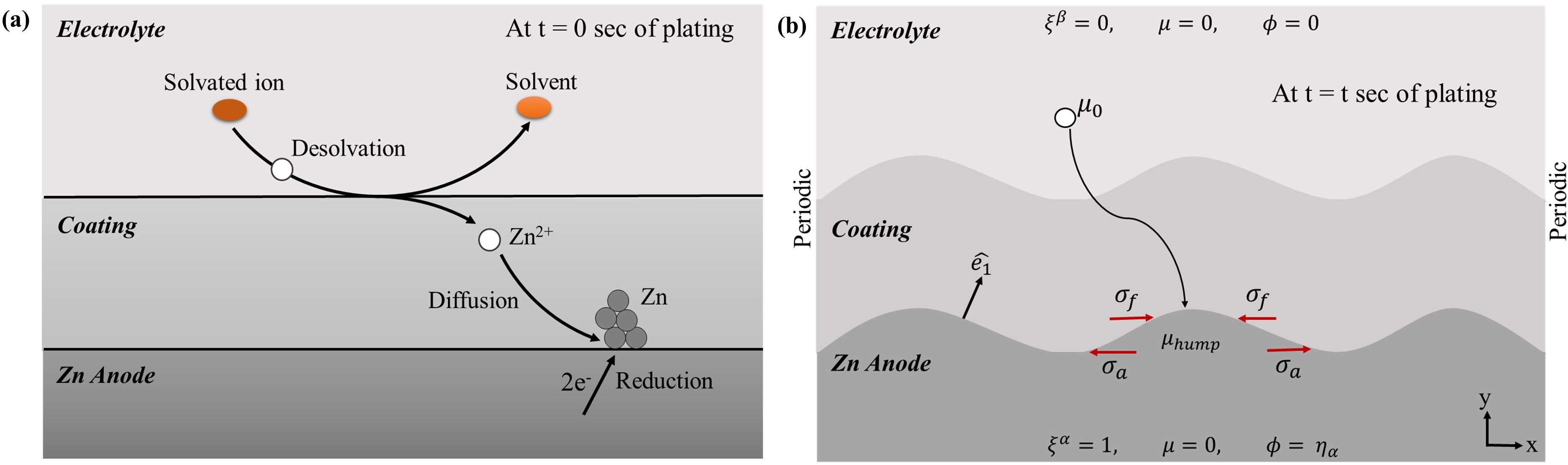}
\qquad
\caption{\label{fig:Schematic_deposition} (a) Schematic diagram illustrating the electrodeposition of zinc ion on the zinc metal anode under the effect of the coating. Here, the dark orange circle represents solvated ion, the light orange circle represents solvent, the dark ash represents metallic Zn, and the white circles represent Zn-ions;
(b) boundary conditions for chemical potential ($\mathrm{\mu}$), electric potential ($\mathrm{\phi}$), and order parameter ($\xi$) associated with the electrochemical phenomena illustrated in (a) along with a schematic of reduction of dendritic growth due to coating due to the impact of thin-film/coating ($\mathrm{\sigma_{f}}$) and anode stress ($\mathrm{\sigma_{a}}$).}
\end{figure}
\section{\label{sec:Results}Results and Discussion}
\par To illustrate the effect of the residual stresses during Zn deposition, we describe the process in the neighborhood of the thin film free-surface, where zinc ions leave their solvent molecules, as shown in~\fig{Schematic_deposition}(a), and settle on the thin film layer's top surface. The Zn ion then diffuses via the thin film layer. When the Zn-ion reaches the Zn metal-thin film interface, it is reduced, and direct deposition on the Zn metal surface is followed by two-electron transfer. Since the interface is flat, there are no preferential sites and deposition happens randomly. However, surface roughness introduces preferred deposition sites via modification of the electric field originating humps or protuberances, ultimately nucleating dendrites (see ~\fig{Schematic_deposition}(b)). Thus, the modified chemical potential ($\mathrm{\mathrm{\mu}}$) along a heteroepitaxial thin film anode interface can be computed knowing the atomic volume $\Omega_0$ as: 
\begin{equation}\label{eq:modified_chemical_potential}
\mu=\mu_0-(\sigma_{\text{x}}+\sigma_{\text{y}}) \Omega_0.
\end{equation}
Here, $\mathrm{\mu_0}$, is the chemical potential of deposited Zn ions, $\mathrm{\sigma_{\text{x}}}$ is the resultant in-plane and $\mathrm{\sigma_{\text{y}}}$ is the stress normal to the anode's surface. The residual stresses ($\mathrm{\sigma_{f}}$, $\mathrm{\sigma_{a}}$) evolving due to to hetero-epitaxy~\cite{10.1016/j.jmps.2024.105897}, contributing to $\mathrm{\sigma_{\text{x}}}$ and $\mathrm{\sigma_{\text{y}}}$,
thus modifying the chemical potential ($\mathrm{\mu}$) and reducing the dendrite growth. Here, we include the effect of the hetero-epitaxial stress in an electro-chemo-mechanical PFM as shown in~\fig{Schematic_deposition}(b) and~\eq{modified_chemical_potential}-  details are included in the methods section in the Supporting Information (SI).

%
%
%

%

%
%

%
\begin{figure}[h]
\centering
{\label{}
\includegraphics[width=0.93\linewidth]{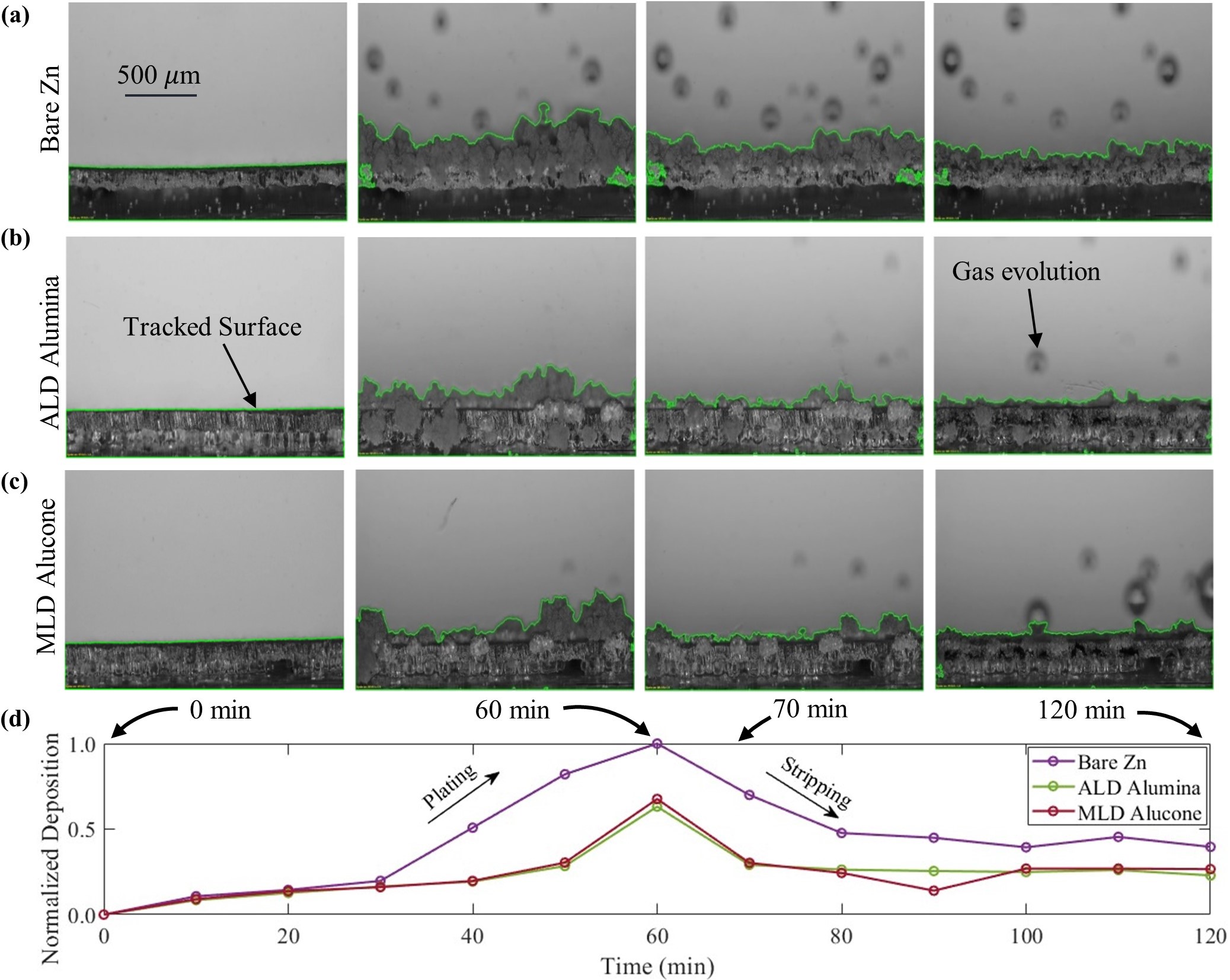}}
\caption{\label{fig:OM_images} Operando imaging of the zinc plating (0-60 min) and stripping (61-120 min) under a current of 8 mA (current density of $\mathrm{10\:mA \cdot cm^{-2}}$) to observe the morphological evolution for (a) bare Zn, (b) one layer (1C) ALD Alumina, and (c) MLD Alucone; (d) Time evolution of the normalized deposition during plating and stripping from pixel counts of OM images (a-c).}
\end{figure}
\par The in-plane interfacial stresses developed by atomic and molecular layered deposition (ALD/MLD) can influence prolonging smooth surface morphology of zinc electrodes, even at high current densities, according to the modified chemical potential shown in~\eq{modified_chemical_potential}. However, dendrites in real life are also influenced by other circumstances, such as charging time, and gas evolution, amongst other factors. Therefore, to investigate the effect of the interfacial stress ($\sigma_\textrm{f}$), we performed charging/discharging of bare Zn and coated thin films at a very high current density (details of the ALD/MLD coatings can be found in the Experimental Methods section of SI). Two different types of coatings were evaluated, including alumina and alucone, with different ALD/MLD cycles (one (1C), five (5C), ten (10C), and fifty (50C) cycles). Higher current density can easily stimulate Zn ion migration, affecting the deposition morphology and the coating effect will be more evident in this case. The dynamic process of Zn electrodeposition and dissolution in aqueous electrolytes was analyzed employing high-throughput operando OM at a consistent current of $\mathrm{10\:mA \cdot cm^{-2}}$, as shown in \fig{OM_images}.~\fig{OM_images} (a) and Movie 1 (see SI) demonstrate the Zn electroplating in the bare $\mathrm{Zn||Zn}$ cell for 60 minutes of plating and 60 minutes of stripping.~\fig{OM_images} (b) and Movie 2, along with~\fig{OM_images} (c) and Movie 3, demonstrate the nucleation and growth of dendrites throughout the plating and stripping on an one layer (1C) alumina and alucone-coated Zn, respectively. 
\par At time 0 minutes, no noticeable features were detected in \fig{OM_images}(a,b,c). The system was in a state of equilibrium as described by the Nernst-Planck equation. A moss-like random dendritic Zn protrusions at 60 min were observed in all coated and uncoated cases, indicating irregular ion migration leads to $\mathrm{Zn^{2+}}$ ion aggregation and nucleation guided by the electric field~\cite{10.1149/1.2411588}. Here, the bare Zn case had higher gas evolution and dendritic growth rate compared to the coated cases, as shown in~\fig{OM_images}. However, gas evolution increased with the increase of coating layers, as shown in~\fig{gas_evolution_charging_discharging}. After 120 mins, due to irregular ion migration and uneven mass transport, the anode surfaces showed residual dendrites.~\fig{OM_images} (d) depicts Zn's normalized deposition (area in each OM frame/maximum area), confirming the fact that the growth rate of the coated surfaces is slower than bare Zn, as the coated deposition is 37\% lower after plating and 13\% lower after stripping. To understand the surface roughness, the surface profiles were extracted and plotted in~\fig{surface_roughness_comparison} (a). Similar to the normalized deposition, the uneven $\mathrm{Zn^{2+}}$ ion aggregation is higher for bare Zn after plating and stripping ends, as shown in~\fig{surface_roughness_comparison} (b). Therefore, the OM experiments verify that the coating, which is as thin as just one ALD/MLD layer (1C), can significantly reduce dendritic growth rates. It is widely recognized that the formation of stresses inside solids can result in changes in their morphology. The equilibrium morphology and morphological stability of strained coating can be determined by balancing the elastic energy with the surface energy~\cite{10.1016/0001-6160(89)90246-0}, and here, the interfacial in-plane stress provides the surface energy to dictate the growth morphology.

%
%

%
\begin{figure}[ht]
\centering
\label{}
\includegraphics[width=1\linewidth]{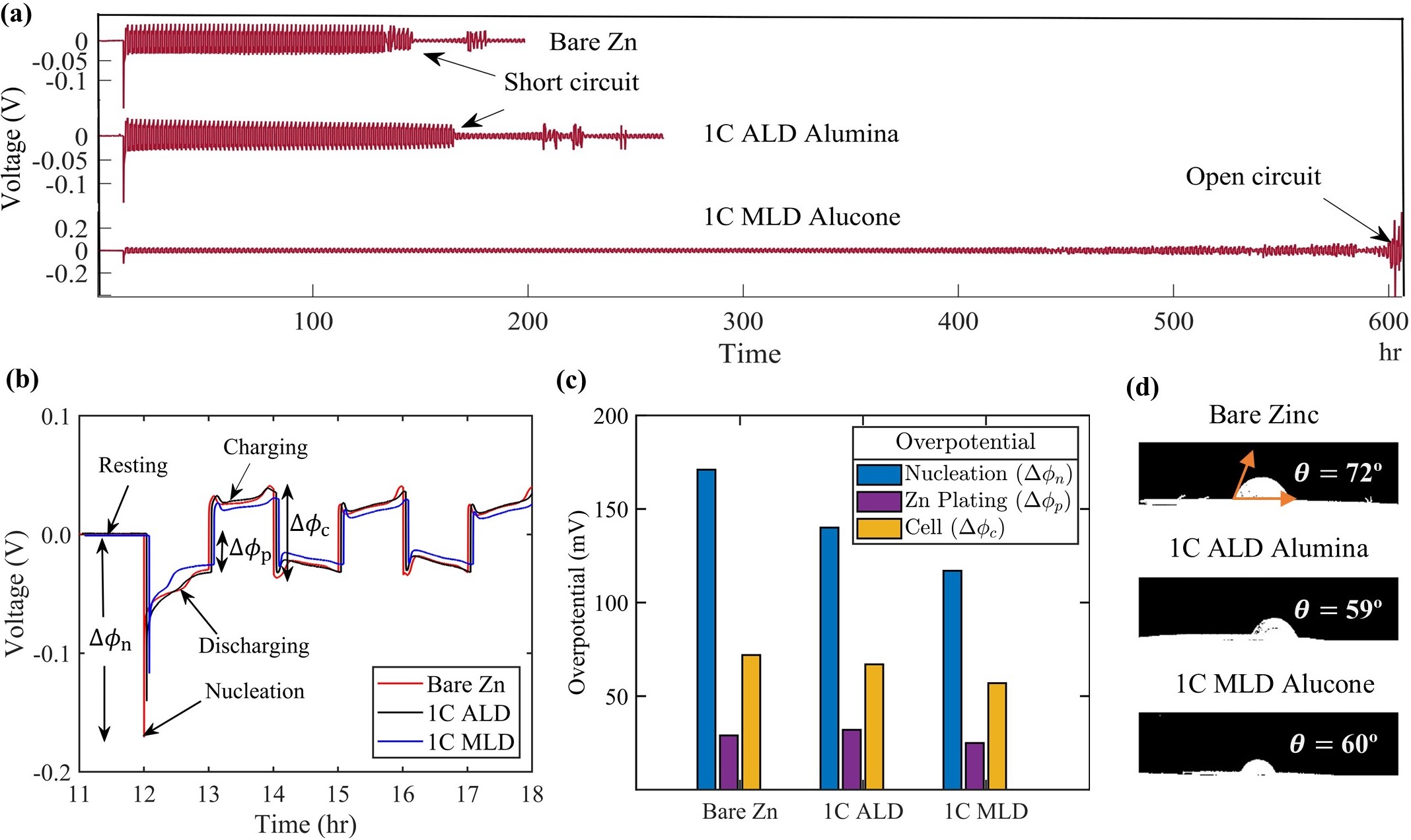}
\caption {(a) Symmetric $\mathrm{Zn||Zn}$ cell charge-discharge profiles with bare Zn, one layer (1C) ALD alumina, one layer (1C) MLD Alucone coating at a current density of $\mathrm{10\: mA \cdot cm^{-2}}$ until short-or open-circuiting at room temperature; (b,c) voltage (millivolts) profiles of first few cycles to observe the patterns of overpotentials; and (d) contact angle measurement for bare and coated Zn foils. } \label{fig:charge_discharge_experiment}
\end{figure}
\par As coating could be a barrier to the diffusion and migration process, we assessed the effect of coating on the cyclic performance of $\mathrm{Zn||Zn}$ cells under a current density of $\mathrm{0.5\:mA \cdot cm^{-2}}$, shown in ~\fig{charge_discharge_experiment}(a). The cycle time of one layer (1C) ALD alumina before short-circuiting or electrolyte depletion (165 hr) is slightly higher than bare Zn (146 hr), whereas 1C MLD alucone carries stable charge-discharge voltage profiles for approximately 598 hr before facing high voltage open-circuit. The results imply that the alucone coating not only reduces the gas evolution and dendrite growth rate but also significantly improves the Zn cyclic performance. Here, for alucone coating, one layer (1C) coating showed the peak cycling life, whereas ten layers (10C) alumina coating showed the peak performance, as shown in~\fig{ALD_MLD_Charge_discharge_cycles_overpotential}. This finding suggests a complex interplay between residual stresses, gas evolution and cycling perfomance in batteries. From~\fig{charge_discharge_experiment}(b), in the first discharge process after 12 hrs of resting, the required nucleation overpotential ($\mathrm{\phi_{n}}$) to begin the Zn plating and stripping is lower for coated Zn compared to bare Zn. The nucleation overpotential for bare Zn, alumina-coated, and alucone-coated cells were -171, -140, and -117 mV, respectively, as shown in~\fig{charge_discharge_experiment}(c). This result indicates that $\mathrm{Zn^{2+}}$ and $\mathrm{e^-}$ transfer resistance decreases with the one layer of coating as lower nucleation potential is required. This is possibly due to the favorable interaction between the coating and electrolyte employed, whereas only one layer of the coating does not create any insulating behavior to increase the overpotential, as mentioned in previous studies~\cite{10.1039/D0TA07232J}. Furthermore, the first plating overpotential ($\mathrm{\phi_{p}}$) to start Zn plating is lowest for alucone-coated Zn (25 mV) compared to bare Zn (29 mV), whereas alumina-coated Zn has the highest (32 mV). On the other hand, cell overpotential ($\mathrm{\phi_{c}}$) shows a similar trend to nucleation overpotential as - bare Zn (72 mV) $\mathrm{>}$ alumina coating (67 mV) $\mathrm{>}$ alucone coating (57 mV). Lower cell overpotential for coated surfaces indicates lower resistance in the batteries, hence better performance for the coated electrodes. The influence of wettability on the ALD alumina and MLD alucone-coated Zn anode was analyzed in a 3 M $\mathrm{Zn(SO_3CF_3)_2}$ aqueous electrolyte by measuring contact angles ($\mathrm{\theta}$) as shown in~\fig{charge_discharge_experiment}(d). The contact angle of alumina-coated Zn ($\mathrm{\theta = 59^{\circ}}$) and alucone-coated Zn ($\mathrm{\theta = 60^{\circ}}$) were much lower than the contact angle of bare Zn ($\mathrm{\theta = 72^{\circ}}$). This suggests that the coated Zn surfaces have improved wettability with the aqueous solution. The enhanced wettability ($\it i.e.,$ lower contact angle, higher adhesion) is advantageous for Zn plating/stripping (cycling) reactions as it can facilitate a consistent flow of Zn ions over the interface between the electrode and electrolyte~\cite{10.1038/s41598-019-51412-5}. This, in turn, reduces the production of Zn dendrites and lowers the charge transfer resistance during battery cycling.
\begin{figure}[ht]
\centering
{\label{}
\includegraphics[width=0.90\linewidth]{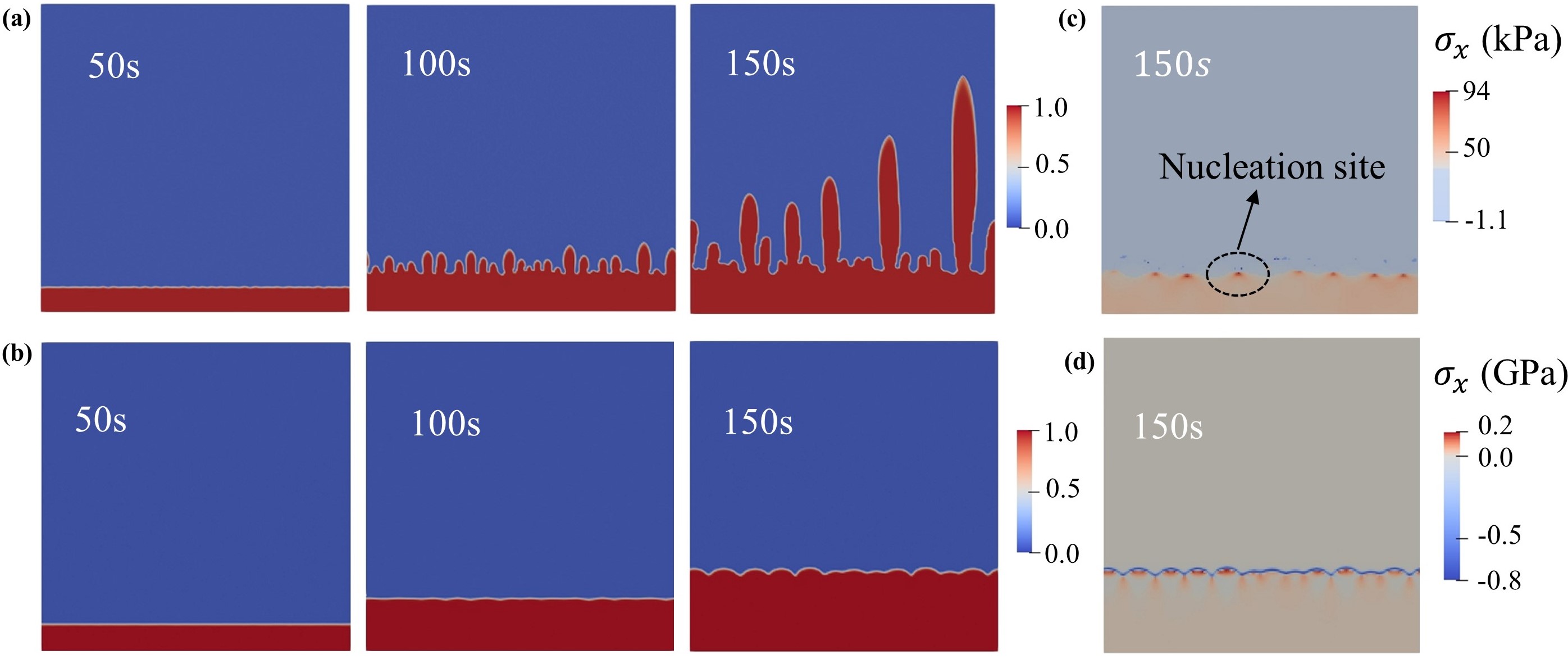}}
\caption{\label{fig:MOOSE_simulations} The development of dendrites ($\mathrm{\xi}$), and the evolution of in-plane stress ($\mathrm{\sigma_x}$) at an applied overpotential ($\mathrm{\phi}$) of 200 mV, current density ($\mathrm{i}$) of $\mathrm{2.8\:mA \cdot cm^{-2}}$, anode stress ($\mathrm{\sigma_{a}}$) of 0.01 GPa, and anisotropy ($\mathrm{\delta}$) of $\mathrm{0.0}$ during 50, 100, and 150 seconds, respectively, for : (a, c) bare Zn, and (b, d) coated  Zn. Here, the coated Zn represents both ALD and MLD coatings from the experiments.}
\end{figure}
\par To demonstrate the impact of coating on dendrite formation using ~\eq{modified_chemical_potential}, we applied the in-plane residual stress as initial interfacial stress, calculated from the heteroepitaxial lattice-misfit for a stable coating on Zn surface~\cite{10.1021/acs.jpcc.2c06646, 10.1016/j.jmps.2024.105897}, modifying the classical Butler-Volmer equation~\cite{10.1016/j.jpowsour.2015.09.055, 10.1021/acsenergylett.8b01009, 10.1021/acsenergylett.0c01235} (See theoretical formulation in SI). Here, the solvated ions gain electrons from the electrode surface during the reduction process (Reduced Zn $\mathrm{\rightleftharpoons}$ Oxidized Zn + $\mathrm{2e^{-}}$) when the overpotential becomes larger than the nucleation energy barrier.
As a result, they overcome the nucleation energy barrier and start depositing on those sites~\cite{10.1149/1.2085653}.  Initial Zn atoms will either freely diffuse to an energy-favorable location or aggregate with other freshly created Zn atoms on the anode surface to generate the initial Zn core~\cite{10.1016/j.mtener.2021.100692}. For the first 50 seconds, there was no undulation on the surface for both bare and coated Zn shown in~\fig{MOOSE_simulations} (a), and~\fig{MOOSE_simulations} (b), respectively. 
At 100 seconds, bare Zn anodes had incipient dendrites formed, whereas, for coated Zn, the interface is pinned but no noticeable dendrites were formed. This suggests that while random dendrite nucleation could happen spontaneously, the residual stresses due to the coating suppresses the uncontrolled dendrite growth from the anode's surface. 
At 150 seconds, the bare Zn surface had prominent dendrites, whereas the coated Zn surface showed a slow dendritic growth rate. One of the reasons behind these inconsistent deposition phenomena is the chemical potential gradient ($\mathrm{\mu}$) as shown in ~\fig{Bare_Zn_MOOSE_thin_film_moose_chemical_potential}. $\mathrm{Zn^{2+}}$ flux shown in~\fig{Bare_Zn_MOOSE_thin_film_moose_chemical_potential} (a) can be explained according to the Nernst-Planck relationship, as $\mathrm{Zn^{2+}}$ diffusion can be caused by the concentration gradient, whereas $\mathrm{Zn^{2+}}$ migration can be caused by an electrostatic potential gradient. The ionic concentration near the anode swiftly drops except for the dendrite tip compared to that in the bulk electrolyte and thus creates a specific concentration gradient as shown in~\fig{Bare_Zn_MOOSE_thin_film_moose_chemical_potential} (a)~\cite{10.1002/cssc.201801657}. 
However, for the coated Zn case, due to the heteroepitaxial in-plane stress (about $\mathrm{\sigma_{f}}$ = 1.2-1.3 GPa for MLD Alucone, see~\fig{ALD_MLD_Charge_discharge_cycles_overpotential}),  the chemical potential polarization is reduced as shown in ~\fig{Bare_Zn_MOOSE_thin_film_moose_chemical_potential} (b).
This causes zinc ions to gather approximately uniformly at the anode surface without favoring dendrite nucleation and growth. To elaborate, the derived interfacial stress has a significant impact on the dispersion of the zinc core's morphology.  The influence of the in-plane GPa range compressive stress shown in  ~\fig{MOOSE_simulations} (d) causes zinc ions close to the zinc anode's surface to adsorb uniformly rather than clustering only in the nucleation sites. Zinc ions deposit readily at the tips of the zinc core, but the compressive stress over the tip of the protuberances is substantially stronger in suppressing the random growth of the Zn cores. On the other hand, as the in-plane stresses are low in the bare Zn surfaces, as shown in~\fig{MOOSE_simulations} (c), abundant charges are gathered over the tips due to the surface's unevenness, where zinc ions are more prone to deposit. Therefore, tips are regarded as active sites for zinc deposition for bare Zn. 
As a result, the comparatively flat electrode surface promotes homogeneous Zn deposition while suppressing Zn dendrite formations, as shown in~\fig{MOOSE_simulations} (d)~\cite{10.1016/j.mtener.2021.100692}.  
\begin{figure}[ht]
\centering
{\label{}
\includegraphics[width=1\linewidth]{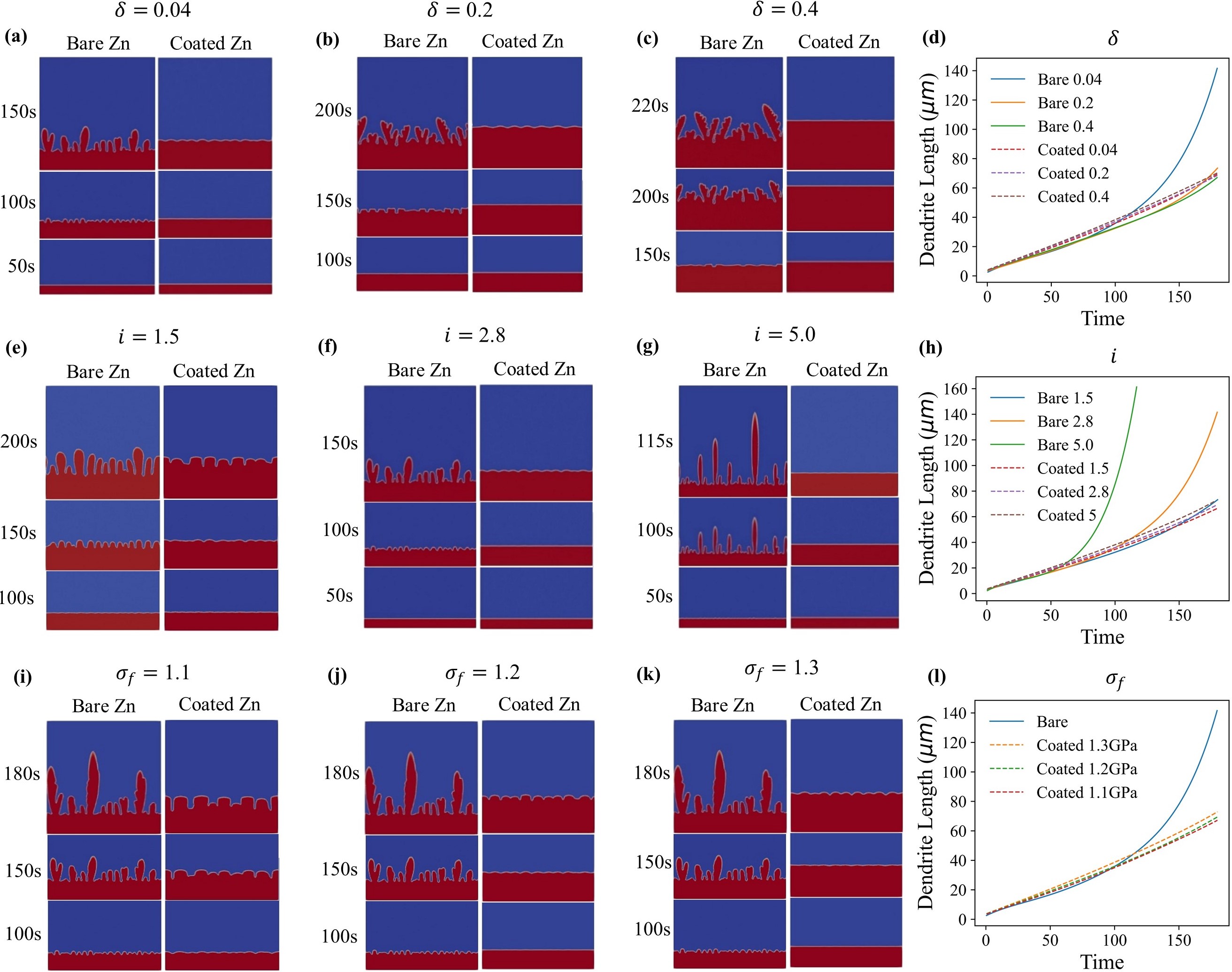}}
\caption{\label{fig:anisotropy_current_density_compare} Morphology evolution for coated and bare Zn under the influence of different conditions: (a,b,c) anisotropy ($\mathrm{\delta= 0.04, 0.2, 0.4}$), (e,f,g) current density ($\mathrm{i= 1.5, 2.8, 5.0 \: mA \cdot cm^{-2}}$), and (i,j,k) in-plane stress ($\mathrm{\sigma_\textrm{f}=1.1, 1.2, 1.3}$ GPa); (d,h,l) average dendrite length with time evolution for all cases.}
\end{figure}
%

%
%
\par To further understand the role of the interfacial stress derived due to coating and the interplay with other cell parameters, we used different material anisotropy and current density to evaluate the effect of the stresses. In~\fig{anisotropy_current_density_compare} (a,b,c), we fixed all parameters except $\delta$ (strength of anisotropy). While increasing $\delta$ gradually from zero (\fig{MOOSE_simulations}(a)) to 0.4 for bare Zn, we observed that the dendrite patterns shift from viscous fingering, $\it{i.e.,}$ perfect isotropic growth ($\delta$ = 0.00, 0.04, ~\fig{anisotropy_current_density_compare} (a)) to side branching ($\delta$ = 0.2, 0.4, see ~\fig{anisotropy_current_density_compare} (b) and (c)). The non-symmetric side branching from principal branches increases with the increase of $\delta$. However, for coated Zn, the growth of the principal dendritic branches starts late as $\delta$ increases, and no substantial dendrites were visible even at 150s, which explains the linear growth of the coated Zn as shown in~\fig{anisotropy_current_density_compare} (d). Therefore, it is clear that initial interfacial stress due to coating is insensitive to $\delta$ and can suppress the dendritic growth.

\par Furthermore, the variation of current density can influence the Zn-ion concentration in the vicinity of the Zn anode shown in
~\fig{anisotropy_current_density_compare} (e,f,g). It is easier for zinc to nucleate unevenly with the high current because there will be a concentration gradient of zinc ions between the bulk solution and the reaction zone~\cite{10.1016/j.mtener.2021.100692}. As a result, for $\mathrm{i=5\: mA \cdot cm^{-2}}$, we saw sharp needle-like dendrites only at 115s, whereas for $\mathrm{i=1.5\: mA \cdot cm^{-2}}$, finger-shaped dendrites started forming at 200s for bare Zn. For all current density variations, the coated Zn surface shows very low undulations on the surface due to the residual stresses. Next, ~\fig{anisotropy_current_density_compare} (i,j,k) shows the sensitivity of dendrite nucleation under variation of residual stress ($\mathrm{\sigma_{f}}$) as there is drastic surface undulation at 150s for $\mathrm{\sigma_{f}}$ = 1.1 GPa, whereas there is comparatively low undulation even at 180s for $\mathrm{\sigma_{f}}$=1.2 and 1.3 GPa. Here, ~\fig{anisotropy_current_density_compare} (d,h,l) indicates the average dendrite length with time and shows the exponential increase of dendrite tip for bare Zn compared to linear growth for coated cases for different anisotropy, current density, and interface stress. Thus, numerical simulations confirmed that coating can reduce the dendrite growth rate in good agreement with in-situ experiments in~\fig{OM_images}. 

\section{\label{sec:Conclusion}Conclusion}
\par By analyzing the bare and coated AZIB, we revealed an in-depth understanding of the influence of in-plane stress on altering chemical potential and, hence, suppressing dendrite growth. 
Firstly, using surface and area tracking of Zn dendrites, we found that subsequent dendrite growth and gas evolution are reduced by coating during Zn electrodeposition, even while applying very high current density. After the first cycle, the bare anode has significantly higher dead Zn and roughness compared to only one layer of ALD/MLD coating.
Secondly, using phase-field modeling, we proved that the interfacial in-plane stresses arising from heteroepitaxial coating suppress the formation of sharp dendrites, whereas bare Zn electrodes show tip splitting, sharp fingers, and branching morphology depending on imposed anisotropy, and current density. 
Thus, our study demonstrates for the first time the effect of residual stresses due to hetero-epitaxial coatings on dendrite suppression and inhibition. 
Critically, we found that these stresses play a critical role in modifying the chemical potential near the surface of the anodes, resulting in a shielding mechanism to suppress dendrite growth. 
Remarkably, the effect of the residual stresses is robust under different current densities and anisotropy of the material, suggesting that the coating strategy of anodes can be an effective method to suppress dendrite growth in AZIB and possibly extendable to all-solid-state batteries \cite{doi:10.1021/acs.jpcc.4c05494}. 

\section{Experimental Methods}
\subsection{Atomic and Molecular Layered Deposited (ALD/MLD) coatings}
Zn foils (0.1 mm in thickness) were sliced into circular discs (diameter, d = 14 mm) and cleansed for 15 minutes by sonication successively in deionized (DI) water and alcohol, followed by natural air drying. Alumina coating on the Zn foils was carried out at $100^{\circ}$C by alternatively providing trimethylaluminum (TMA) and $\mathrm{H_2O}$, whereas alucone coating was performed by alternatively providing trimethylaluminum (TMA) and ethylene glycol ($\mathrm{HO-CH_2-CH_2-OH}$) into a commercial GEMStar\texttrademark ~XT Atomic Layer Deposition System. The actuator keeps open to increase the pressure in the chamber and purges everything. When the actuator is closed, the chamber pressure stays at operating pressure (500-700 mTorr) when pulsing happens. The sequence for TMA pulsing is: 200 ms pulse of TMA, 5 s delay, delay 15 s (actuator opened), actuator closed. The sequence for the 2nd precursor ($\mathrm{H_2O}$/EG) pulsing is: 21 ms pulse of $\mathrm{H_2O}$/EG, 5 s delay, delay 15 s (actuator opened). Alumina and alucone coatings were applied to Zn foils with just 1 ALD/MLD (1C) cycle to get one layer of coating. Similarly, 5, 10, and 50 cycles (C) of the coating were applied for both alumina and alucone.
%
\subsection{Battery Assembly, and Electrodeposition}
Symmetric $\mathrm{Zn\vert\vert Zn}$ coin cells were fabricated using Zn metal discs (d = 14 mm, thickness ($\mathrm{h_s}$) 0.1 mm), separators (Whatman, glass fiber, GF/F, diameter d=19 mm), and 3 M Zn trifluoromethanesulfonate ($\mathrm{Zn(SO_3CF_3)_2}$) aqueous electrolyte (75 $\mathrm{\mu}$L). The coin cells were crimped with a hydraulic crimping machine at 1000 psi pressure. Galvanostatic charge/discharge tests for these two coating materials were conducted in a Neware battery testing system (Neware 4000). These tests, including all coated (1C, 5C, 10C, 50C) and bare Zn cases, were carried out at a fixed current density ($\mathrm{j}$) of 0.5 $\mathrm{mA \cdot cm^{-2}}$ to analyze the cycling performance of AZIBs. All batteries were tested at room temperature ($25~^{\circ}$C).
\subsection{Operando Optical Microscopy and Batch Image Processing}
Using an optical microscope (Olympus BX53M, Japan), in-situ monitoring of symmetric Zn$\vert\vert$Zn cell's dendrite evolution and dissolution was carried out. A $5\times$ objective lens was employed to achieve a natural aperture of 0.15 and a resolution of about 3.35 $\mu$m. Electrochemical tests with high constant current density modes (10 mA$\cdot$cm$^{-2}$, 8 mA over 8$\times$10 mm electrode) were synchronized with picture collection. Digital photographs were taken constantly each minute for the first two to five cycles to record the dynamic growth until the screen got blurred. The electrochemical data is stored every 4 seconds. Subsequently, the captured images were converted from RGB to 8-bit images. To facilitate faster image processing to analyze the dynamic growth, datasets were batch-processed and tracked employing an open-source software named SurfTrack, using a color threshold between dendrites and background after binarization. The dendrite area and Zn dendrite height (surface roughness) were also quantified using SurfTrack.
%

\begin{suppinfo}
This material is available free of charge via the Internet at https://pubs.acs.org.
\begin{itemize}
    \item  \sloppy List of symbols, Theoretical Formulation of the Phase-Field Method, Computational Implementation; Results: In-situ optical microscopy for multiple coated layers, Cyclic test for multiple coated layers, and Chemical potential evolution.
    \item \sloppy Dendrite growth under in-situ optical microscopy for Bare Zn, 1C ALD Alumina and 1C MLD Alucone are added as movie, namely, $\mathrm{movie1\_Bare\_Zn.mp4}$, $\mathrm{movie2\_1C\_ALD\_Alumina.mp4}$, and $\mathrm{movie3\_1C\_MLD\_Alucone.mp4}$, respectively.

\end{itemize}
\end{suppinfo}

\section{\label{sec:Code_Availability}Code Availability}
\begin{itemize}
  \item All post-processing of in-situ optical microscopy images to capture dendrite length and normalized deposition is done using the “SurfTrack” python package built available at: \sloppy \href{https://github.com/MusannaGalib/SurfTrack} {https://github.com/MusannaGalib/SurfTrack}.
  \item All post-processing of MOOSE simulation's exodus file format is done using the “MOOSEanalyze” python package built on Paraview's PvPython and available at: \sloppy \href{https://github.com/MusannaGalib/MOOSEanalyze} {https://github.com/MusannaGalib/MOOSEanalyze}.
\end{itemize}

\section{\label{sec:Code_Availability}Data Availability}
All cycling data (Bare Zn, 1C ALD Alumina, 1C MLD Alucone) is available at: \sloppy \href{https://github.com/MusannaGalib/MOOSEanalyze/tree/main/CyclicTest} {\sloppy \\ https://github.com/MusannaGalib/MOOSEanalyze/tree/main/CyclicTest}.

\section{\label{sec:Author_Contributions}Author Contributions}
\textbf{Musanna Galib:} Conceptualization, Investigation, Software, Formal analysis, Data Curation, Validation, Writing - Original Draft, Writing - Review \& Editing.~\textbf{Amardeep Amardeep:} Investigation, Writing - Review \& Editing.~\textbf{Jian Liu:} Investigation, Writing - Review \& Editing, Supervision, Funding acquisition.~\textbf{Mauricio Ponga:} Conceptualization, Investigation, Writing - Review \& Editing, Supervision, Project administration, Funding acquisition.

\section{\label{sec:ACKNOWLEDGMENTS}Acknowledgement}
We acknowledge the support from the New Frontiers in Research Fund (NFRFE-2019-01095), the Discovery grant from the Natural Sciences and Engineering Research Council of Canada (NSERC) under Award Application Number 2016-06114, Collaborative Research Mobility Award (UBC CRMA) and the UBC Eminence program (Battery Innovation Cluster). M.G. gratefully acknowledges the financial support from the Department of Mechanical Engineering at UBC through the Four Years Fellowship. This research was supported through high-performance computational resources and services provided by Advanced Research Computing at the University of British Columbia. M.G. would also like to express gratitude to Biswajeet Rath for the helpful discussions about phase-field modeling and Zhenrui Wu for the helpful discussions about electrochemical experiments.
\bibliography{main.bib}

\clearpage
\raggedbottom
\setcounter{section}{0}
\setcounter{equation}{0}
\setcounter{figure}{0}
\setcounter{table}{0}
\setcounter{page}{1}
\makeatletter
\renewcommand{\thesection}{S\arabic{section}}
\renewcommand{\theequation}{S\arabic{equation}}
\renewcommand{\thefigure}{S\arabic{figure}}
\renewcommand{\thetable}{S\arabic{table}}
\renewcommand{\thepage}{SM\arabic{page}}

\renewcommand{\bibnumfmt}[1]{[S#1]}
\renewcommand{\citenumfont}[1]{S#1}

\begin{center}
{\Large \bf Supporting Information}
\end{center}
\subsection{List of symbols}
\tab{variables_symbols} outlines all the parameters applied in the current phase-field model discussed in the Theoretical Method section.

\subsection{Theoretical Formulation of the Phase-Field Method}

\begin{table}[h!]
\centering
\caption{Physical parameters (variables/constants) and their symbols}
\begin{tabular}{||c|c||c|c||}
\hline
{Parameter Name} & {Symbol} & {Parameter Name} & {Symbol} \\
\hline \hline
Order parameter & $\xi$ & Electric potential & $\phi$ \\
\makecell{Local electrochemical free \\energy density} & $f_{\text {elec-ch}}$ & Interfacial energy density & $f_\textrm{in}$ \\
Elastic energy density & $f_{\text {ela}}$ & Double well function & $W$ \\
Gradient energy coefficient & $\kappa$ & Langevin noise & $f_\textrm{noise}$ \\
Anisotropic modulus & $\omega$ & Dendrite angle & $\theta$ \\
Heteroepitaxial compressive stress & $\sigma_\text{f}$ & In-plane stress & $\sigma_{\text{film}}$ \\
Interfacial stress & $\sigma_\text{x}$ & Normal stress & $\sigma_\text{y}$ \\
Modified chemical potential & $\mu$ &  \makecell{Chemical potential of \\ deposited Zn} & $\mu_{0}$ \\
\makecell{Chemical potential of dendritic\\ hump} & $\mu_{\text{hump}}$ &  \makecell{Chemical potential of \\ unperturbed system} & $\mu^*$ \\
Activation overpotential & $\eta_{\mathrm{\alpha}}$ & \makecell{Standard equilibrium half\\ cell potential} & $E^{\theta}$ \\
Interfacial mobility & $L_\sigma$ & Exchange current density & $i_0$ \\
\makecell{Electro-Chemical kinetic\\ coefficient} & $L_\eta$ & Electrons transferred & $n$ \\
Molar volume & $V_m$ & Surface tension & $\gamma$ \\
Gas constant & $\text{R}$ & Charge transfer coefficient & $\alpha$ \\
Evolution time & $t$ & Temperature & $\text{T}$ \\
\makecell{Molar fraction of Zn ion} & $c_{\textrm{Zn}^{2+}}$ & \makecell{Molar fraction of Zn\\ ion at t=0} & $c_0$ \\
\makecell{Molar fraction of Zn\\ in electrode} & $c^\text{l}$ & \makecell{Molar fraction of Zn\\ in electrolyte} & $c^\text{s}$ \\
Site density of electrode & $C_\text{m}^\text{s}$ & Site density of electrolyte & $C_\text{m}^\text{l}$ \\
Conductivity of electrode & $\sigma^\text{s}$ & Conductivity of electrolyte & $\sigma^\text{l}$ \\
Atomic volume & $\Omega_0$ & Interpolation function & $h(\xi)$ \\
Zn electrode stiffness & $C_{\text{Zn}}$ & Diffusivity of Zinc ion & $\mathrm{D}$ \\
\hline
\end{tabular}
\label{tab:variables_symbols}
\end{table}

%


\par The electrodeposition mechanism in $\mathrm{Zn}$ batteries is described by the electrochemical-mechanical phase field model (PFM). 
The $\mathrm{Zn}^{2+}$ ions transfer from the electrolyte to the zinc anode due to the electrochemical potential difference is replicated in this model. 
The Butler-Volmer equation is used to simulate how the accumulated Zn affects the evolution of the Zn-thin film interface (\textit{i.e.}, moving upward) as electrodeposition proceeds~\cite{10.1039/C8TA07997H}.
However, to unveil the full potential of PFM, decoupling of interfacial energy and interfacial thickness is highly desirable to achieve a sufficiently good resolution of the interface~\cite{10.1103/PhysRevE.98.023309}, which is a limitation for Wheeler, Boettinger, and McFadden's~\cite{10.1103/PhysRevA.45.7424} pioneering alloy solidification model and Kim, Kim, and Suzuki's phase-field framework~\cite{10.1103/PhysRevE.60.7186}.
Here, we used an alloy solidification model developed by Plapp~\cite{10.1103/PhysRevE.84.031601}, based on a grand-potential functional that preserves the benefit of decoupling interfacial thickness from interfacial energy while eliminating the necessity for variables of phase concentration in the chemical free energy term.   
Instead of composition, an evolution equation based on the difference in chemical potential among species is utilized, which has been used in several previous studies~\cite{10.1103/PhysRevE.85.021602, 10.1016/j.actamat.2015.03.051, 10.1103/PhysRevE.95.063312}.
The free energy functional for an electrochemical interface can be presented as:~\cite{10.1149/1945-7111/ac22c7}
  \begin{equation} \label{eq:gibbs_free_energy}
  \begin{split} 
f\left(\xi, c_{\text{Zn}^{2+}}, \phi\right)=\int_{V}[f_{\text {elec-ch }}\left(\xi, c_{\textrm{Zn}^{2+}}, \phi\right)+f_{\text {in }}(\xi)
+f_{\text {ela }}(\xi)] \mathrm{d} V,
   \end{split}
   \end{equation} 
where $\xi$ describes the transition from the electrolyte phase ($\xi$ = 0) to the $\mathrm{Zn}$ solid phase ($\xi$ = 1) and represents a non-conserved order parameter (phase-field variable);  $\phi$ and $c_{\textrm{Zn}^{2+}}$ denote the electric potential and the $\mathrm{Zn}^{2+}$cation concentration, respectively; $f_{\text {elec-ch}}\left(\xi, c_{\textrm{Zn}^{2+}}, \phi\right)$, $f_\text{in}(\xi)$, and $f_{\text {ela }}(\xi)$ are the local electrochemical free energy density, the interfacial energy density, and elastic energy density, respectively.
\par In the phase field model, the Zn anode's Ginzburg-Landau type
 heterogeneous interface energy density has the following form: 
  \begin{equation} \label{eq:heterogeneous_interface_energy_density}
  \begin{split} 
f_{in}=W(\xi)+\frac{1}{2} \kappa(\nabla \xi)^{2}+f_{\text {noise}}.
   \end{split}
   \end{equation} 
The double well function, $W(\xi)=W_{0}(1-\xi)^{2} \xi^{2}$ in~\eq{heterogeneous_interface_energy_density}, characterizes the equilibrium states by defining the bulk energy needed to phase change from electrode to electrolyte (here, $\mathrm{W_{0}=2}$~\cite{10.1016/j.jpowsour.2015.09.055}). Let $\kappa$ be a gradient energy coefficient and $\frac{1}{2} \kappa(\nabla \xi)^{2}$ is the gradient energy density that correlates to the interface strength. $f_{\text {noise}}$ measures the amount of interfacial heterogeneous deposition brought on by defects. On the anode interface, conserved Langevin noise is used to prevent concentration drift in simulations that ensure mass conservation.  Langevin noise, $f_{\text {noise}}$, is incorporated into ~\eq{heterogeneous_interface_energy_density} to consider the disturbance in the system caused by surface imperfections and thermal deviations that may initiate the development of the dendrite nucleus.
\par The gradient energy coefficient's anisotropy is expressed using Kobayashi's proposed formula as~\cite{10.1016/0167-2789(93)90120-P} 
$\kappa=\kappa_{0}[1+\delta \cos (\omega \theta)].$
Here, $\omega$ is the anisotropic modulus which defines the number of preferential growth orientations of dendrites ($w = 6$ for hcp Zn~\cite{10.1103/PhysRevE.92.011301}), $\theta$ is the angle between the crystallographic direction and the surface normal vector, and $\delta$ is the energy anisotropy coefficient/strength of anisotropy of hexagonal close-packed Zn crystal. It is a microscopic interaction width that affects the interface's movement to add anisotropy by considering that $\delta$ relies on the orientation of the interface's outer normal vector. 
\par To characterize the stress effect during plating, the elastic energy density is introduced: 
  \begin{equation} \label{eq:elastic_energy_density}
  \begin{split} 
f_{\text {ela }}(\xi)=\frac{1}{2} C_{\textrm{Zn}}\boldsymbol{\varepsilon}_{\textrm{Zn}}{ }^{2},
   \end{split}
   \end{equation} 
where the effective stiffness of zinc solid is represented by $C_{\textrm{Zn}}$ (assumed isotropic in this case). For calculating $f_{\text {ela }}$, the equilibrium equation of solid is solved to depict the stress-strain behavior by solving for displacement field $\boldsymbol{u}$ in the domain $\Omega_A$, which has the following strong form:
$\nabla \cdot \boldsymbol{\sigma}+\boldsymbol{b}=0 \textrm{\:\:in\:} \Omega_A$, where $\boldsymbol{\sigma}$ denotes the Cauchy stress tensor and $\boldsymbol{b}$ denotes the body force. The finite strain increment, total strain, and incremental rotation are computed for the generic time ($t$) increment such that $t \in\left[t_n, t_{n+1}\right]$. The resulting strain is calculated referencing the deformed arrangement ($\textit{i.e.,}$ Eulerian definition) using the Eulerian-Almansi finite strain tensor as $\boldsymbol{\varepsilon}_{\textrm{Zn}} = \frac{1}{2} (I - \boldsymbol{\hat{C}^{-1} })$. Here, $\boldsymbol{\hat{C}}= \boldsymbol{\hat{F}} \boldsymbol{\hat{F}}^{T}$ is incremental right Cauchy-Green deformation tensor, $\hat{\boldsymbol{F}}=\frac{\partial \boldsymbol{x}_{n+1}}{\partial \boldsymbol{x}_n} = \boldsymbol{F}_{n+1} \boldsymbol{F}_n^{-1}$; $\hat{\boldsymbol{F}}$ is incremental deformation gradient, and $\boldsymbol{F_n}$ is the total deformation gradient at time $t_n$; $\boldsymbol{x}$ is the position vector ($\boldsymbol{x_{n+1}} = \boldsymbol{x_n} + \boldsymbol{u}$).
\par 
Developed heteroepitaxial residual compressive stresses ($\sigma_{\text{f}}$)~\cite{10.1016/j.jmps.2024.105897} in the thin film are implemented leveraging the diffused interface between the two phases in phase field model as the interface stress as shown in~\fig{Schematic_deposition} (b) of the manuscript and defined by the gradient of an order parameter as $\mathrm{\boldsymbol{\sigma}_{\text{film}} \cdot \textit{f}(\mid\nabla\xi\mid)}$. Therefore, the 2D stress tensor,  $\boldsymbol{\sigma}_{\text{film}} = \left(\boldsymbol{Q} \cdot \boldsymbol{M} \cdot \boldsymbol{Q}^{-1}\right)$, and $f(|\nabla \xi|) = |\nabla \xi|$, where  $\boldsymbol{M}=\left(\begin{array}{ccc}
0 & 0  \\ 0 & \sigma_f \end{array}\right),
\boldsymbol{Q}=(\begin{array}{ll} \vec{e}_1 & \vec{e}_2 \end{array})$. Here,  $\boldsymbol{Q}$ and $\boldsymbol{Q}^{-1}$ are applied as the basis transformation from the eigenvector basis to the cartesian basis. $\vec{e}_1=\frac{\nabla \xi}{|\nabla \xi|}$ eigenvector is applied in the direction of the order parameter gradient as shown in ~\fig{Schematic_deposition}(b), with an eigenvalue of zero, whereas $\vec{e}_2$ is perpendicular to that direction. 
\par 

%
\begin{figure}[h]
\centering
\label{}
\includegraphics[width=0.45\linewidth]{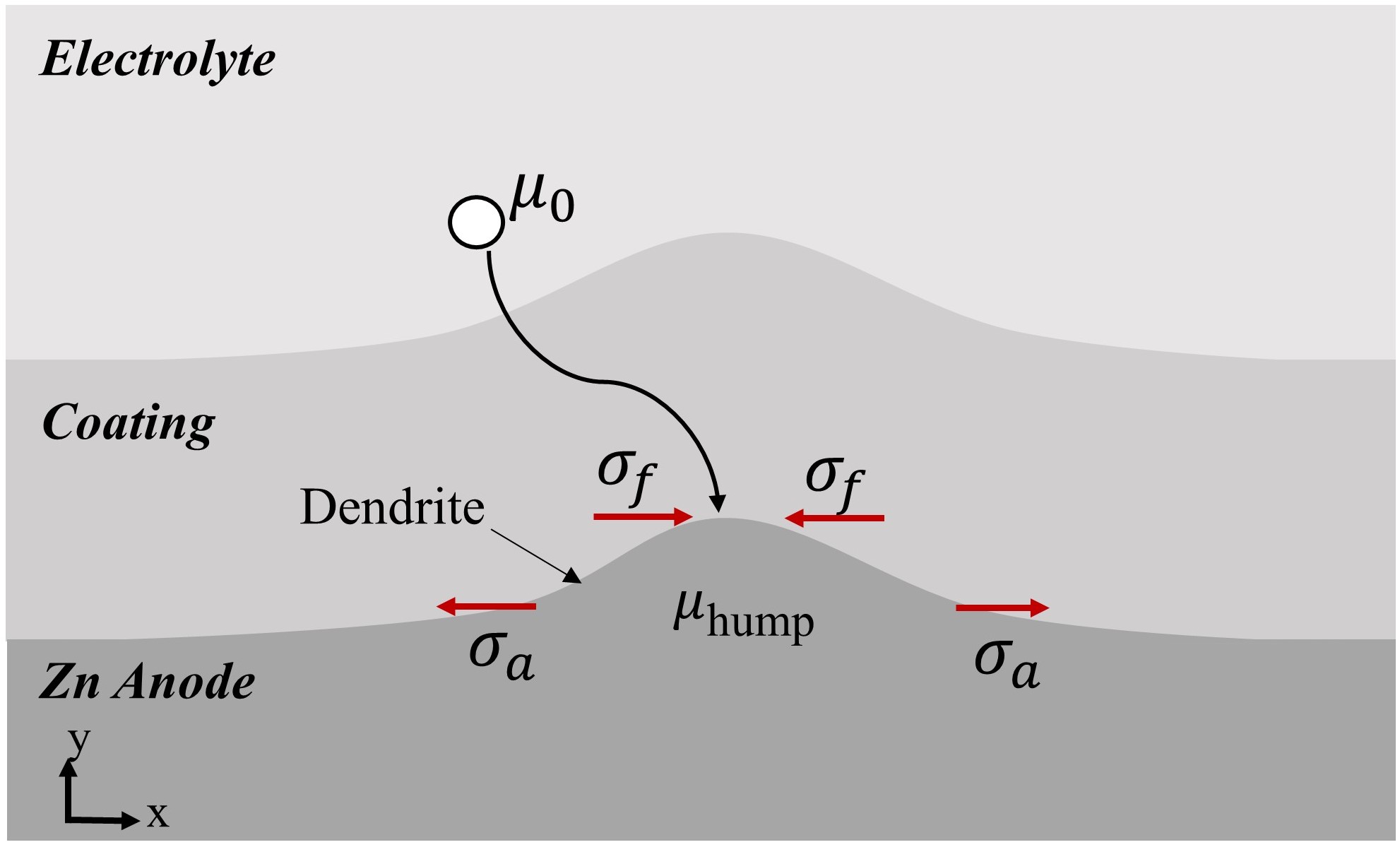}
\caption{\label{fig:schematic_chemical_potential} Scheamitc diagram illustrating impact of thin-film/coating ($\mathrm{\sigma_{f}}$) and anode stress ($\mathrm{\sigma_{a}}$) on chemical potential. Relative sizes not at scale in the schema.}
\end{figure}
From~\fig{schematic_chemical_potential}, assuming the humps form during plating in the anode with an addition of $\delta N$ atoms with $\delta v$ volume change to the Zn crystal, the work done to add the new atoms will be $\mathrm{\left(\mu^*-\mu_{\textrm{hump}}-\mu_{0}\right)  \delta} N$; where $\mu^*$, $\mu_{\textrm{hump}}$ and $\mu_{0}$ are chemical potential of the unperturbed system, dendritic hump and deposited Zn ions, respectively. If the free energy of the hump is $\delta F$, we can write the work done as $\mathrm{\left(\delta \textit{F}-\mu_{0} \delta \textit{N}-\sigma_{\text{x}} \delta v-\sigma_{\text{y}} \delta v\right)}$, where $\sigma_{\text{x}}$ and $\sigma_{\text{y}}$ are interfacial (due to hetero-epitaxy)~\cite{10.1016/j.jmps.2024.105897} and normal stresses on the surface. Here, $\mu_{0} \delta N$ denotes the Zn ions bulk reference energy, and we assume $\delta F=\mu_{0} \delta N$  for a stable hump without any spontaneous desolvation. Therefore, by equating these two works, we get
$
\mathrm{\mu^*-\mu_{\text{hump}}=\mu_{0}-\sigma_{\text{x}} \Omega_{0}-\sigma_{\text{y}} \Omega_{0}}~\cite{10.1063/1.1699681},
$
where $\mathrm{\Omega_{0}=\delta v / \delta }N$ is the atomic volume. Therefore, the modified chemical potential along a heteroepitaxial thin film anode interface, $\mathrm{\mu}$, is:
\begin{equation}\label{eq:modified_chemical_potential}
\mu=\mu_0-\sigma_{\text{x}}\Omega_0-\sigma_{\text{y}} \Omega_0.
\end{equation}
Here, the interfacial stress ($\mathrm{\sigma_{\text{x}}}$) is related to interfacial tension as $-\mathrm{\sigma_{\text{x}}=\gamma \Omega_0 \textit{K}(x)}$~\cite{10.1016/0001-6160(89)90246-0}. Additionally, $\mathrm{\gamma}$ is the interfacial tension, $\mathrm{\Omega_0}$ denotes the atomic volume, and $\mathrm{\textit{K}(x)}$ refers to the curvature of the interface. The second term on the right-hand side of~\eq{modified_chemical_potential} reflects the contribution of surface energy to the chemical potential. The third component quantifies the impact of stress, which is perpendicular to the interfacial contact, on the release or absorption of an atom at that interface.

%
\par Therefore, to add the effect of modified chemical potential in the electrochemical free energy density ($\mathrm{f_{\text {elec-ch}}\left(\xi, c_{\text{Zn}^{2+}}, \phi\right)}$), the activation over-potential ($\mathrm{\eta_{\alpha}}$) has to be modified. $\eta_{\alpha}$ can be written as $\eta_{\alpha} = \phi - E^{\theta}$, $E^{\theta}$ is the standard equilibrium half-cell potential of the electrode relative to the electrolyte, and $\phi=\phi^{\text {etrode }}-\phi^{\text {elyte }}$ is the applied over-potential. Here, $\phi^{\text {etrode }}$ is the potential in the electrode neighboring to the interfacial region, and $\phi^{\text {elyte }}$ is the electrolyte potential adjacent to that interface. Ganser \emph{et al.} presented the equilibrium cell potential as~\cite{10.1149/2.0221906jes, 10.1149/2.1111904jes}
\begin{equation}\label{eq:modified_equilibrium_cellpotential}
 \begin{aligned}
    \textrm{F}E^{\theta} &= \mu^{\text {elyte }}-\mu^{\text {etrode }} \\
    &= \mu^{\text {elyte }}_0-\mu^{\text {etrode }}_0 -\sigma_{\text{x}}\Omega_0-\sigma_{\text{y}} \Omega_0 \\
    &= \textrm{F}E^{\theta}_0 -\sigma_{\text{x}}\Omega_0-\sigma_{\text{y}} \Omega_0,
 \end{aligned}
\end{equation}
where $\mu^{\text {etrode}}$ refers to the chemical potential of the Zn metal ion specifically in the electrode, and $\mu^{elyte}$ is in the electrolyte. Here,  F denotes the Faraday constant $\left(\textrm{F} = 96485\: \mathrm{C} \cdot \mathrm{mol}^{-1}\right)$ and $\mu^{\text {elyte }} = \mu^{\text {elyte }}_0$; neglecting the contribution of stress to the
the chemical potential of the Zn ion in the electrolyte. These chemical potentials can be controlled by stress as shown in ~\eq{modified_chemical_potential}, just as they can be by any other non-ideality or departure from a datum state. Using~\eq{modified_equilibrium_cellpotential}, we can write the $\eta_{\alpha}$ as
\begin{equation}\label{eq:modified_overpotential}
\eta_{\alpha}=\left(\phi^{\text {etrode }}-\phi^{\text {elyte }}-E^{\theta}_0\right)+ (\sigma_{\text{x}}+\sigma_{\text{y}})\frac{\Omega_{0}}{\textrm{F}}.
\end{equation}
\par For the electrochemical reaction in the anode ($\textrm{Zn}^{2+}+2e^{-} \rightleftharpoons \textrm{Zn}$), the spatial evolution of free energy density $(f_{\text {elec-ch}}\left(\xi, c_{\text{Zn}^{2+}}, \phi\right))$ with respect to $\xi$ ($\frac{\partial f_{\text {elec-ch}}}{\partial \xi}$) coupled with the solid and ions phases of Zn and the overpotential $\eta_{\alpha}$ takes the following shape  
\begin{equation} \label{eq:electrochemical_free_energy_density}
  \begin{split} 
\begin{aligned}
\Gamma &=\frac{\partial f_{\text {elec-ch}}}{\partial \xi} \\
&= \Biggl\{\exp\Biggr[\frac{(1-\alpha)n\textrm{F}\eta_{\alpha}}{\textrm{RT}}\Biggr]-\frac{c_{\textrm{Zn}^{2+}}}{c_{o}}\exp\Biggr[\frac{-\alpha n\textrm{F}\eta_{\alpha}}{\textrm{RT}}\Biggr]\Biggl\}\\
&= \Biggl\{\exp\Biggr[\frac{(1-\alpha)n\textrm{F}\left(\phi-E^{\theta}_0\right)+(1-\alpha) \Omega_{0} (\sigma_{\text{x}} +\sigma_{\text{y}})}{\textrm{RT}}\Biggr]\\
&-\frac{c_{\textrm{Zn}^{2+}}}{c_{0}}\exp\Biggr[\frac{-\alpha n\textrm{F}\left(\phi-E^{\theta}_0\right)-\alpha \Omega_{0} ( \sigma_{\text{x}}+ \sigma_{\text{y}}) }{\textrm{RT}}\Biggr]\Biggl\}.\\
\end{aligned}
   \end{split}
   \end{equation} 
In the electrodeposition case, under high overpotential when the cathodic reaction current is much smaller than the anodic reaction current,~\eq{electrochemical_free_energy_density} can be simplified by neglecting the mechanical contribution to the cathodic reaction current,
\begin{equation} \label{eq:electrochemical_free_energy_density_simplified}
  \begin{split} 
\begin{aligned}
\Gamma
= \Biggl\{\exp\Biggr[\frac{(1-\alpha)n\textrm{F}\eta_{\alpha}}{\textrm{RT}}\Biggr]
-\frac{c_{\textrm{Zn}^{2+}}}{c_{o}}\exp\Biggr[\frac{-\alpha n\textrm{F}\eta_{\alpha}-\alpha  \Omega_{0}( \sigma_{x} + \sigma_{y} )}{\textrm{RT}}\Biggr]\Biggl\}.\\
\end{aligned}
   \end{split}
   \end{equation} 
\par The temporal growth of the phase-field variable $\xi$ in~\eq{gibbs_free_energy} using classical Butler-Volmer equation can be expressed as \cite{10.1016/j.jpowsour.2015.09.055, 10.1021/acsenergylett.8b01009, 10.1021/acsenergylett.0c01235}
  \begin{equation} \label{eq:order_parameter}
  \begin{split} 
  \frac{\partial\xi}{\partial t}&=-L\frac{\partial f\left(\xi, c_{\text{Zn}^{2+}}, \phi\right)}{\partial\xi}\\&= -L\left[\frac{\partial f_{\text {in }}(\xi)+ \partial f_{\text{ela}}(\xi)}{\partial \xi}+\Gamma\right]+f_{\text {noise}}\\
&=-L_{\sigma}(W^{\prime}(\xi)-\kappa\nabla^{2}\xi) - L_{\sigma} \frac{\partial f_{\text{ela}}(\xi)}{\partial \xi}
    -L_{\eta}h^{\prime}(\xi)\Biggl\{\exp\Biggr[\frac{(1-\alpha)n\textrm{F}\eta_{\alpha}}{\textrm{RT}}\Biggr]\\&-\frac{c_{\textrm{Zn}^{2+}}}{c_{0}}\exp\Biggr[\frac{-\alpha n\textrm{F}\eta_{\alpha}}{\textrm{RT}}\Biggr]\exp\Biggr[\frac{-\alpha(\sigma_{\text{x}} +\sigma_{\text{y}}) \Omega_{0}}{\textrm{RT}}\Biggr]\Biggl\}+f_{\text {noise}}.
   \end{split}
   \end{equation} 
Here, $L_{\sigma}$ refers to the interfacial mobility activities ($2.5 \times 10^{-7}$ m$^3\cdot$J$^{-1} \cdot$s$^{-1}$)~\cite{10.1021/acsenergylett.0c01235} and $L_{\eta}=\frac{V_{\mathrm{m}} \gamma}{F \kappa} i_{0}$ refers to the electro-chemical kinetic coefficient. 
From~\eq{order_parameter}, $V_{\mathrm{m}}$, $i_{0}$, $\gamma$, $\alpha$, n, R, t, and T are the molar volume of $\mathrm{Zn}$ ($\mathrm{9.16\:cm^3}$)~\cite{10.1103/PhysRevE.92.011301}, the exchange current density (varies from $\mathrm{1.5-5\: mA}\cdot\mathrm{cm^{-2}}$ in this study), surface tension ($\mathrm{0.5\: J \cdot m^{-2}}$)~\cite{10.1103/PhysRevE.92.011301}, charge-transfer coefficient (fixed to $0.5$ in this study), number of electrons transferred  (2 for $\mathrm{Zn}$ electrodeposition), gas constant$\left(\text{R} = 8.314 \mathrm{~J} \cdot \mathrm{mol}^{-1} \cdot \mathrm{K}^{-1}\right)$, evolution time, and temperature ($\mathrm{300}$ K), respectively. Note that the value for Langevin noise ($f_{\text{noise}}$) was fixed to $0.04$ for this study.  
The interfaces between the electrode and electrolyte have a finite thickness, where the value of $\xi$ lies between 0 and 1. 
Here, a polynomial tilting/interpolation function $h(\xi) = \xi^{3}(6\xi^{2} - 15\xi + 10)$~\cite{10.1016/j.jpowsour.2015.09.055} is used for smoothly interpolating different physical parameters between two phases. Its derivative, $h^{\prime}(\xi)=30 \xi^{2}(1-\xi)^{2}$, is only valid at the interface as $h^{\prime}$ becomes zero when $\xi$ proceed towards 0 or 1. Therefore, using $h^{\prime}(\xi)$ for interpolation of physical parameters ensures that the electrochemical reactions occur solely at the interface.
The variables $c_{\text{Zn}^{2+}}$ and $c_{0}$ represent the molar ratios of zinc ions at a particular location and at the beginning (at t = 0 s, $\frac{1}{c_{0}}=58.556$), respectively. Here, $c_{\textrm{Zn}^{2+}}$ can be expressed as~\cite{10.1103/PhysRevE.92.011301}
  \begin{equation} \label{eq:molar_concentration}
  \begin{split} 
  c_{\textrm{Zn}^{2+}} = c^{(\text{l,s})}(1-h(\xi))=\frac{\exp\bigr[\frac{(\mu_0-\epsilon^{(\text{l,s})})}{\textrm{RT}}\bigr]}{1+\exp\bigr[\frac{(\mu_0-\epsilon^{(\text{l,s})})}{\textrm{RT}}\bigr]}(1-h(\xi)),
   \end{split}
   \end{equation} 
The variables $c^{\text{l}}$ and $c^{\text{s}}$ represent the molar fraction of Zn in the electrolyte and electrode phase, respectively. The term $\epsilon^{(\text{l,s})} = \mu_0^{(\text{l}_0,\text{s}_0)} - \mu_0^{\text{N}_0}$ refers to the difference in the chemical potential of Zn and neutral elements in the electrolyte at the initial equilibrium condition.
%
%
%
%
%
\par The chemical potential ($\mu_0$) of the Zn ions can be derived from the mass conservation law, as~\cite{10.1021/acsenergylett.0c01235}:
  \begin{equation} \label{eq:rewrite_flux_concentration_and_mass_conservation}
  \begin{split} 
\chi \frac{\partial \mu_0}{\partial t}=\nabla \cdot \frac{D c_{\textrm{Zn}^{2+}}}{\textrm{RT}}[\nabla \mu_0+n \textrm{F} \nabla \phi]-\frac{\partial h(\xi)}{\partial t}\left[c^{\text{s}} \frac{C_{\text{m}}^{\text{s}}}{C_{\text{m}}^{\text{l}}}-c^{\text{l}}\right],
   \end{split}
   \end{equation} 
where the susceptibility factor $\chi=\frac{\partial c^{\text{l}}}{\partial \mu_0}[1-h(\xi)]+\frac{\partial c^{\text{s}}}{\partial \mu_0} h(\xi) \frac{C_{\text{m}}^{\text{s}}}{C_{\text{m}}^{\text{l}}}$ and $\text{D}$ is the diffusivity ($\mathrm{3.68 \times 10^{-10}\:m^2\cdot s^{-1}}$)~\cite{10.1103/PhysRevE.92.011301}. Therefore, $\mu_0$ can be determined from the revised diffusion equation~\eq{rewrite_flux_concentration_and_mass_conservation} where $C_{\text{m}}^{\text{s}}$ ($\mathrm{1.092 \times 10^5\: mol \cdot m^{-3}}$)~\cite{10.1103/PhysRevE.92.011301} and $C_{\text{m}}^{\text{l}}$ ($\mathrm{5.652 \times 10^4\: mol \cdot m^{-3}}$)~\cite{10.1021/acsenergylett.0c01235} represent the electrode and electrolyte site densities which are inverse of molar volume. 
%
The distribution of $\phi$ over the spatial domain can be acquired from the effective conductivity ($\sigma^\text{c}$), i.e., from conduction equation~\cite{10.1021/acsenergylett.8b01009}
  \begin{equation} \label{eq:conduction_equation}
  \begin{split} 
  \nabla\sigma^\text{c}\nabla\phi = n\textrm{F}C_{\text{m}}^{\text{s}}\frac{\delta(\xi)}{\delta t}.
   \end{split}
   \end{equation} 
Here, $\sigma^\text{c}$ is determined by the conductivity of the electrode phase, $\sigma^{\text{s}}$ ($\mathrm{10^7\:S \cdot m^{-1}}$)\cite{10.1021/acsenergylett.0c01235}, and the conductivity of the electrolyte phase, $\sigma^{\text{l}}$ ($\mathrm{4.64\:S\cdot m^{-1}}$ for 1 M of $\mathrm{ZnSO_4}$)~\cite{10.1021/je0600911} and can be expressed as ${\sigma^\text{c} = \sigma^{\text{s}}h(\xi) + \sigma^{\text{l}}[1 - h(\xi)]}$.\\

\subsection{Computational Implementation}
\begin{figure}[h]
\centering
\label{}
\includegraphics[width=0.45\linewidth]{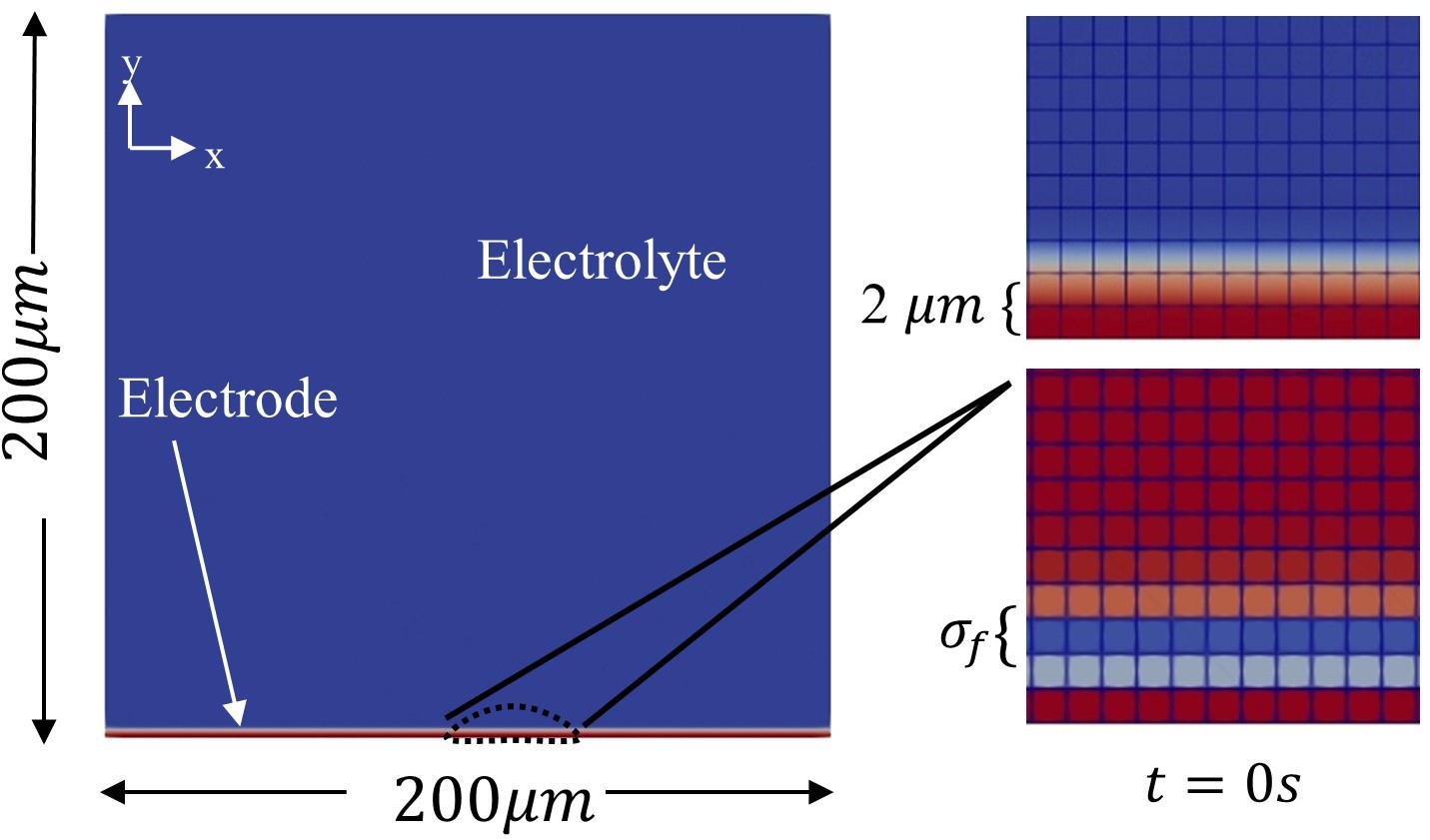}
\caption{\label{fig:simulation_cell} Diagram illustrating the simulated cell's initial frame prior to electrodeposition and the mesh size.}
\end{figure}
\par In this study, we utilized the open-source MOOSE software~\cite{10.1016/j.softx.2022.101202, 10.1002/nme.5284} to develop the grand potential-based nonlinear phase-field model. This model investigated a set of coupled nonlinear equations ($\mathrm{\textit{i.e.}}$, phase-field equation coupled with chemical potential and conduction equations) to understand the effect of the thin film-deposited residual stress on electrodeposition kinetics. The phase-field model computes the changes across time and space of three independent variables ($\mathrm{\xi, \mu, \phi}$). In MOOSE,~\eq{order_parameter},~\eq{molar_concentration},~\eq{rewrite_flux_concentration_and_mass_conservation} and~\eq{conduction_equation} are converted to its weak form and solved simultaneously by minimizing the total residual. The coupled variables were normalized to ensure that they weighed almost equally. Here,  the coupling between the variables is as follows: the phase-field variable $\mathrm{\xi}$ of~\eq{order_parameter} relies on variable $\mathrm{\mu}$ via the local zinc-ion molar fraction using~~\eq{molar_concentration}. Secondly, based on \eq{rewrite_flux_concentration_and_mass_conservation}, it can be deduced that $\mathrm{\mu}$ is influenced by $\mathrm{\phi}$ from  \eq{conduction_equation} and $\mathrm{\xi}$ from \eq{order_parameter}, whereas $\mathrm{\phi}$ in~\eq{conduction_equation} only depends on $\mathrm{\xi}$. To maintain the accuracy of the boundary conditions, the cutoff distance is established at a value that is less than 90\% of the simulation size. The initial geometry before starting the first plating at t = 0 s displayed in~\fig{simulation_cell} is illustrated by the distribution of $\mathrm{\xi}$ in the following equation
%
  \begin{equation} \label{eq:distribution_of_order_parameter}
  \begin{split} 
  \xi(x, y)=\frac{1-\tanh [2(x-2)]}{2},
   \end{split}
   \end{equation} 
which denotes the anode thickness of $\mathrm{2 ~\mu m}$ with a smooth shift at the interface from the anode to the electrolyte. The remaining $\mathrm{198 ~\mu m}$ is occupied by the electrolyte in a $\mathrm{200 ~\mu m \times 200 ~\mu m}$ domain as shown in~\fig{simulation_cell}. The electric potential $\mathrm{\phi}$ is initially specified as
$
\mathrm{\phi(x, y)=\phi_{\text {applied }} \xi(x, y)}
$
such that the anode and electrolyte have the potential of $\mathrm{\phi_{\text {applied }}}$ and 0, respectively, at $t = 0$ s. $\phi_\textrm{applied}$ represents the overpotential applied over the interface. The chemical potential ($\mu$) was also set to zero at t = 0 s for the whole domain. 
The spatial distribution of all independent variables ($\mathrm{\xi, \mu, \phi}$) are resolved using Newton-Raphson's iterative approach at $0.01 \mathrm{~s}$ time step along with adaptive time steps. Here, we normalized the parameters (time, length,  temperature, and stress) using the following normalization factors: $ 1 \mathrm{~s}, 1~\mu \mathrm{m}, 1 ~\mathrm{K}$, and $\mathrm{1 \times 10^{9} \mathrm{~Pa}}$, respectively. All post-processing and analysis of the MOOSE simulations has been performed using the open-source Python package called "MOOSEanalyze".


\subsection{\label{sec:Introduction}Results}

\subsubsection{\label{sec:Introduction}In-situ optical microscopy for multiple coated layers}

\begin{figure}[H]
\centering
{\label{}
\includegraphics[width=0.95\linewidth]{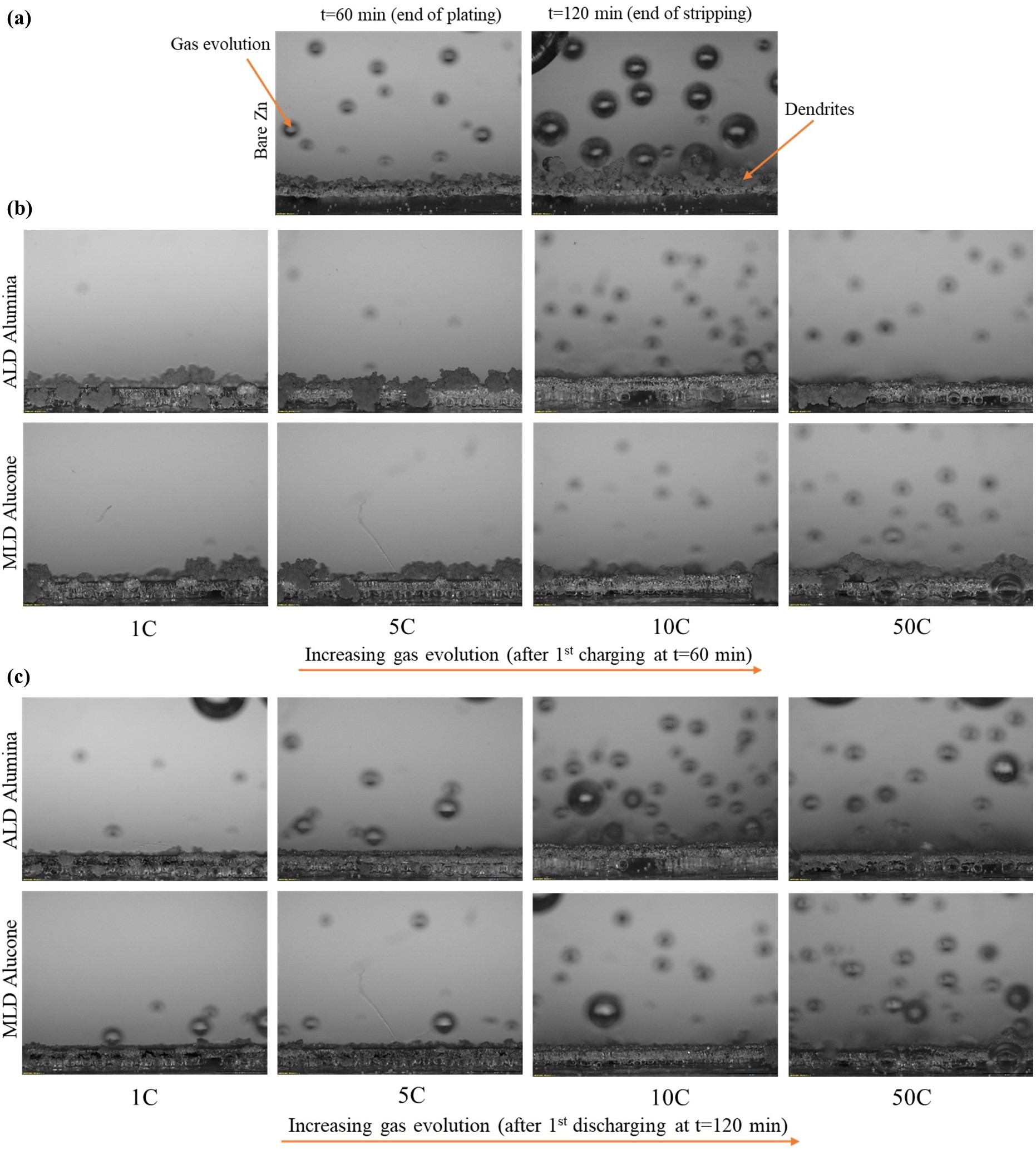}}
\qquad
\caption{\label{fig:gas_evolution_charging_discharging} Gas evolution observation from in-situ optical microscopy at $\mathrm{i=10~mA \cdot cm^{-2}}$  (a) bare Zn, (b) charging states for coated Zn at $\mathrm{t=60}$ min, and (c) discharging stages for coated Zn at $\mathrm{t=120}$ min.}
\end{figure}
\begin{figure}[h]
\centering
{\label{}
\includegraphics[width=0.99\linewidth]{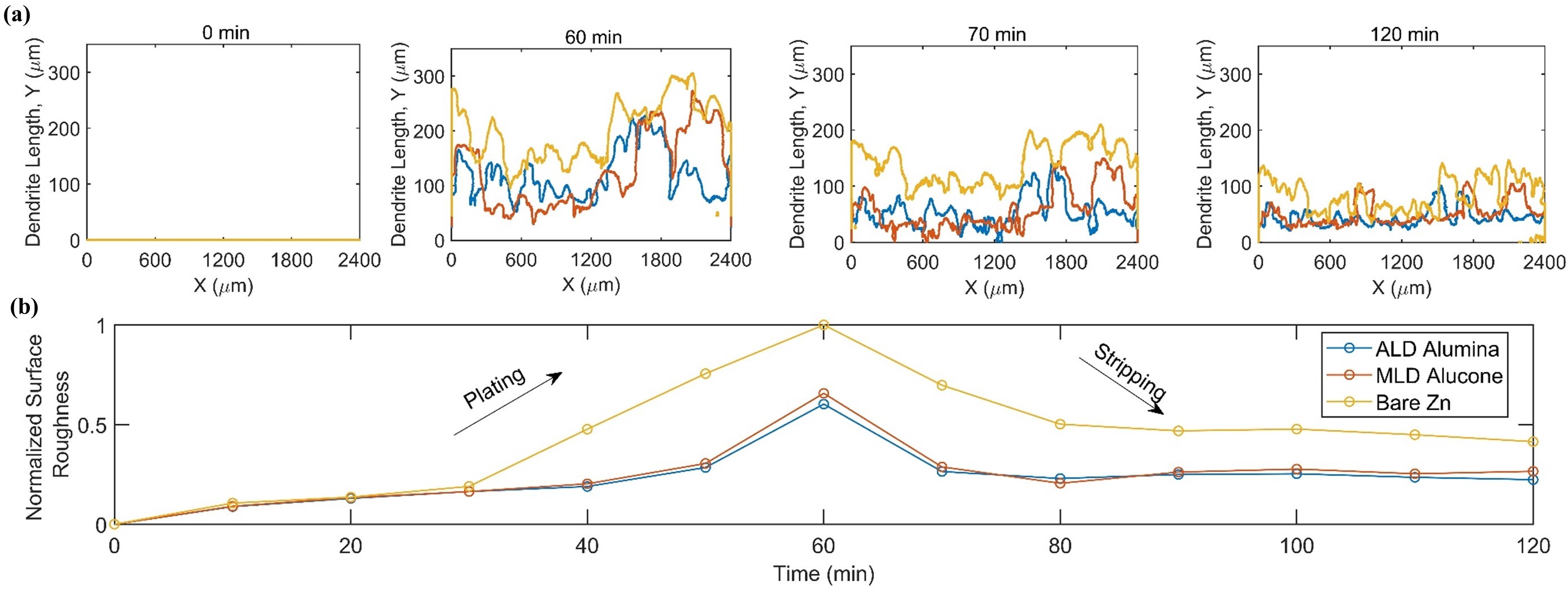}}
\caption{\label{fig:surface_roughness_comparison} (a) dendrite profile extracted during plating (0,~60 min) and stripping (70,~120 min); and (b) normalized surface roughness with time for coated and bare Zn.}
\end{figure}
\subsubsection{\label{sec:Introduction}Cyclic test for multiple coated layers}

\begin{figure}[H]
\centering
{{
\label{}
\includegraphics[width=0.99\linewidth]{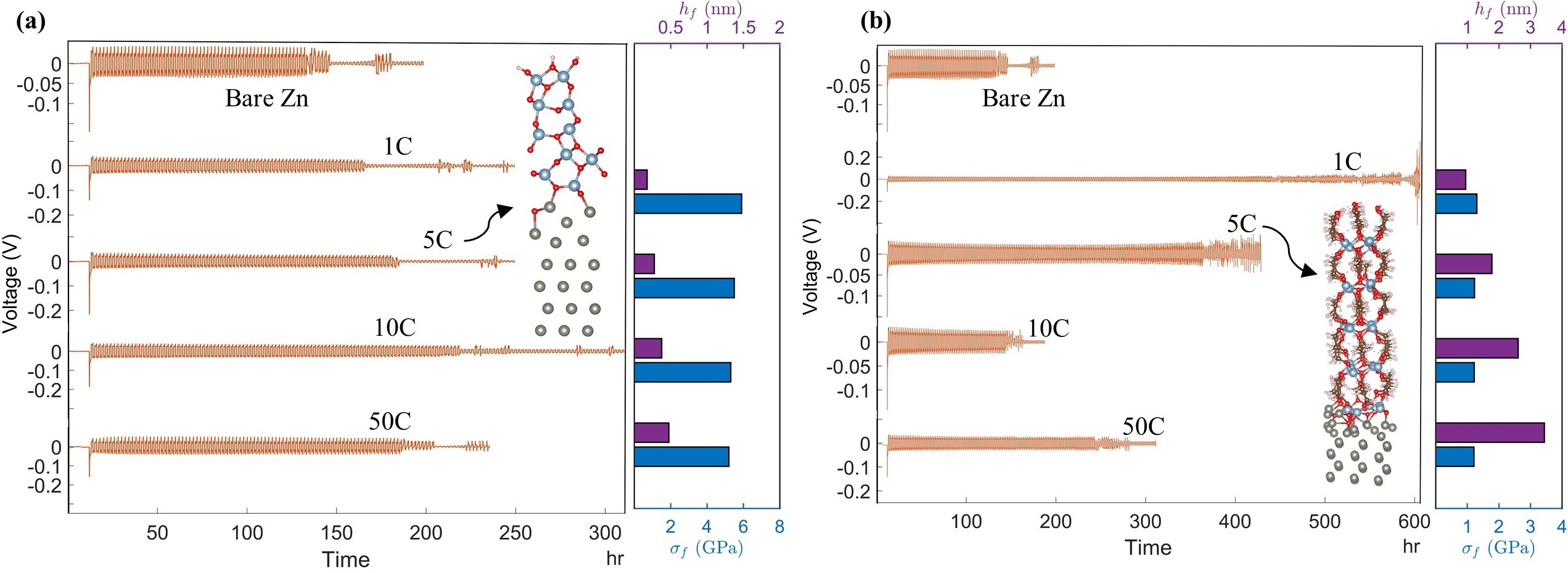}}}
\caption{\label{fig:ALD_MLD_Charge_discharge_cycles_overpotential} Charging/discharging cycles for Bare and coated Zn at different coating thicknesses (1C, 5C, 10C, 50C) with their corresponding hetero-epitaxial stress for (a) ALD Alumina, (b) MLD Alucone. Color code for the atoms in 5C alumina \& alucone: carbon - brown, zinc - gray, aluminium - blue, hydrogen - pink, and oxygen - red.}
\end{figure}

In~\fig{ALD_MLD_Charge_discharge_cycles_overpotential}, one (1C), five (5C), ten (10C), and fifty (50C) coated layers of ALD alumina and MLD alucone were cycled to understand the effect of coating thickness on cyclic life. Here, the coating thickness (${h_\text{f}}$) and lattice misfit ($\varepsilon_\text{m}$) were measured from density functional theory (DFT) optimized structures~\cite{10.1021/acs.jpcc.2c06646} as shown in the inset of  ~\fig{ALD_MLD_Charge_discharge_cycles_overpotential}(a), (b) for 5C case. Then, the coating generated hetero-epitaxial stress is calculated using the analytical formulation:
$  
\sigma_{\textrm{f}} = \frac{\varepsilon_\textrm{m}}{\left(\frac{1}{M_{}^\textrm{T}}+\frac{4h_\textrm{f}}{h_{\textrm{s}}M_{}}\right)}~\cite{10.1016/j.jmps.2024.105897}
$  using their stiffness properties (M)~\cite{10.1021/acs.jpcc.2c06646} and Zn foil thickness ($\mathrm{h_s}$ = 0.1mm). The range of interfacial compressive stress is found to be 5.2-5.9 GPa for ALD Alumina and 1.2-1.3 GPa for MLD Alucone, which is the applied initial interfacial stress in our phase-field formulation.
\subsubsection{\label{sec:Chemical_potential_evolution}Chemical potential evolution}
\begin{figure}[ht]
\centering
{\label{}
\includegraphics[width=0.95\linewidth]{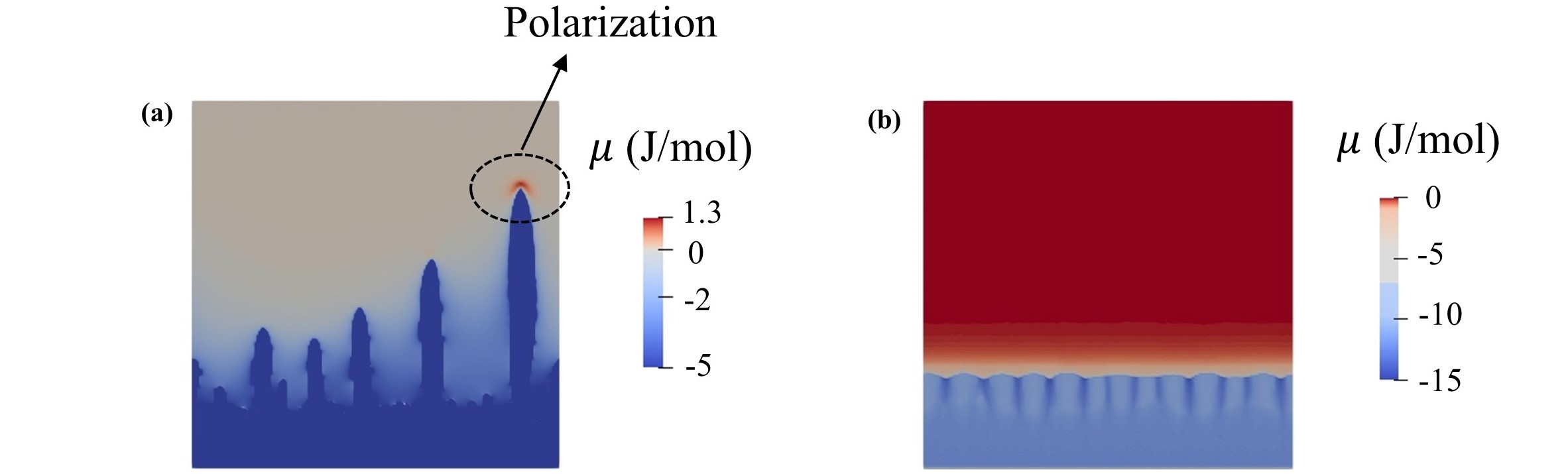}}
\caption{\label{fig:Bare_Zn_MOOSE_thin_film_moose_chemical_potential} The change in chemical potential ($\mathrm{\mu}$) at a applied overpotential ($\mathrm{\phi}$) of 200 mV, current density ($\mathrm{i}$) of ($\mathrm{2.8\:mA\cdot cm^{-2}}$) and anisotropy ($\mathrm{\delta}$) of $\mathrm{0.0}$ at 150 seconds, respectively, for : (a) bare Zn, and (b) coated  Zn.}
\end{figure}

\end{document}